\documentclass[12pt,a4paper]{article}
\usepackage{a4wide}
\usepackage{latexsym}
\usepackage{epsfig}
\usepackage{amssymb}
\usepackage{eufrak}

\def\be{\begin{equation}}
\def\ee{\end{equation}}
\def\bea{\begin{eqnarray}}
\def\eea{\end{eqnarray}}
\def\bc{\begin{center}}
\def\ec{\end{center}}
\def\bb{\begin{thebibliography}{99}}
\def\eb{\end{thebibliography}}
\def\pd{\partial}
\def\a{\alpha}
\def\b{\beta}
\def\g{\gamma}
\def\d{\delta}
\def\m{\mu}
\def\n{\nu}

\def\l{\lambda}
\def\s{\sigma}
\def\e{\epsilon}
\def\p{\pi}

\def\RR{\mathbb{R}}
\def\CC{\mathbb{C}}
\def\ZZ{\mathbb{Z}}
\def\HS{{\cal{H}}}
\def\FF{{\cal{F}}}

\def\KK{{\cal{K}}}
\def\LL{{\cal{L}}}
\def\MM{{\cal{M}}}

\makeindex

\begin{document}

\newcommand{\newc}{\newcommand}
\newc{\sm}{Standard Model}

%%%%%%%%%%%%%%%%%%%%%%%%%%%%%%%%%%%%%%%%%%%%%%%%%%%%%%%%%%%%%%%%%%%%%%%%%%%%%%%%%%%%%%%%%%%%%%%%%%%%%%%%%%%%%%%%%%%%%%
%%%%%%%%%%%%%%%%%%%%%%%%%%%%%%%%%%%%%%%%%%%%%%%%%%%%%%%%%%%%%%%%%%%%%%%%%%%%%%%%%%%%%%%%%%%%%%%%%%%%%%%%%%%%%%%%%%%%%%
\thispagestyle{empty}
\linespread{1}

\begin{flushright}
IFT-UAM/CSIC-04-08\\
hep-th/0405074\\
\end{flushright}

\vspace{1cm}

\begin{center}

{\bf\Large Topics on D-brane charges with B-fields}

\vspace{.7cm}

{\bf Juan Jos\'e Manjar\'{\i}n}\footnote{E-mail: {\tt juanjose.manjarin@uam.es}} \\

\vspace{1cm}

{\it   
 Instituto de F\'{\i}sica Te\'orica, C-XVI,
  Universidad Aut\'onoma de Madrid \\
  E-28049-Madrid, Spain}

\begin{abstract}

In this review we show how K-theory classifies RR-charges in type II string theory and how the inclusion of the B-field modifies the general structure leading to the twisted K-groups. Our main purpose is to give an expository account of the physical relevance of K-theory and, in order to make it, we consider different points of view: processes of tachyon condensation, cancellation of global anomalies and gauge fixings. As a field to test the proposals of K-theory, we concentrate on the study of the D6-brane, now seen as a non-abelian monopole.

\end{abstract}
\end{center}
%%%%%%%%%%%%%%%%%%%%%%%%%%%%%%%%%%%%%%%%%%%%%%%%%%%%%%%%%%%%%%%%%%%%%%%%%%%%%%%%%%%%%%%%%%%%%%%%%%%%%%%%%%%%%%%%%%%%%%
%%%%%%%%%%%%%%%%%%%%%%%%%%%%%%%%%%%%%%%%%%%%%%%%%%%%%%%%%%%%%%%%%%%%%%%%%%%%%%%%%%%%%%%%%%%%%%%%%%%%%%%%%%%%%%%%%%%%%%
\baselineskip = 15pt
\tableofcontents
%%%%%%%%%%%%%%%%%%%%%%%%%%%%%%%%%%%%%%%%%%%%%%%%%%%%%%%%%%%%%%%%%%%%%%%%%%%%%%%%%%%%%%%%%%%%%%%%%%%%%%%%%%%%%%%%%%%%%%
%%%%%%%%%%%%%%%%%%%%%%%%%%%%%%%%%%%%%%%%%%%%%%%%%%%%%%%%%%%%%%%%%%%%%%%%%%%%%%%%%%%%%%%%%%%%%%%%%%%%%%%%%%%%%%%%%%%%%%
\section{Introduction}\label{Intro}

D-branes play a central role in String Theory since they allow us to test the web of dualities that relate the different string theories and, therefore, to find out which is the vacuum manifold.

Besides the intuitive idea that D-branes are the hyperplanes where open strings can end, there is a large amount of unsolved questions concerning them. One of the major improvements in this subject was the discovery that D-branes are BPS states that carry RR-charges, \cite{Pol}. These charges are described in terms of $p$-forms potentials $C_p$ for $p$ taking odd values in type IIA and even values in type IIB and, as diferential forms they were studied in cohomological terms.

However, more recent research has revealed that the two defining properties of D-branes, stated in the previous paragraph, should be handled with care. On one hand, Sen \cite{Sen.2} has shown that the BPS condition can be relaxed, \cite{Sen.1, Taq}. The original Sen's conjecture involved the bosonic string and his proposal is that the open string lives in the background of an unstable D25-filling brane and the tachyonic mode is such that

\begin{itemize}

\item[$i$)] its potential has a locally stable minimum, whose energy density measured with respect to that of the unstable point is equal to minus the tension of the D25-filling brane;

\item[$ii$)] the lower dimensional D-branes are solitonic solutions of the theory in the background of this D25-brane;

\item[$iii$)] the locally stable minimum corresponds to the closed string vacuum and, therefore, there are no open string excitations.

\end{itemize}

These criteria can be easily adapted to the supersymmetric case. For type IIA we can consider a system of unstable D9-filling branes \cite{Hor} which, due to the action of the GSO projection, do not carry any conserved RR-charge and have a tachyonic instability. For type IIB, D9-filling branes carry a RR-charge and form a doublet of $SL(2,\ZZ)$ with the NS9-branes. However, there is bound state which do have a tachyonic instability, and it is the D9-$\bar {\mbox{D}}$9 system, where the GSO projection does not cancel the tachyons coming from the open strings with one end in the D9-brane and the other in the $\bar {\mbox{D}}$9-brane.

However, there is a difference between the bosonic and supersymmetric constructions, and it is the fact that, in the supersymmetric case, these unstable states do not neccesarily deacy to the closed string vacuum, but they can carry lower dimensional charges, corresponding to different stable states of the theory. With this construction one can recover the whole spectrum of both theories and obtain some new results, such as stable but nonsupersymmetric states which are stable since they are the lowest states of the theory and cannot further decay \cite{Sen.2}. 

On the other hand, although the RR potentials are differential forms, they are not classified by cohomology. Instead, RR-charges (and fluxes) arise as characteristic classes in K-theory \cite{Min.Moo, Wit.1, Moo.Wit}, a sort of generalized cohomology theory that establishes equivalence relations, not between differential forms, but between fiber bundles.

Intuitively this can be understood from the fact that these RR fields in the bulk couple to the Wess-Zumino (WZ) terms appearing in the effective world-volume theory of the D-branes. Therefore, RR fields retain some information on the gauge bundles, the Chan-Paton bundles, living in the world-volume and it is precisely this information which is classified in terms of K-theory.

In the case of type II string theories the Chan-Paton degrees of freedom furnish representations of the unitary group $U(N)$ \cite{Wit.2}, where $N$ depends on the number of D-branes lying at the same location of space-time, producing an enhancement of symmetry. This gives rise to the principal or vector bundles classified by K-theory.

Let us give a technical result, \cite{Min.Moo}, that shows that the anomalous couplings on the world-volume of D-branes, \cite{Min.Moo, Gree.Har.Moo, Che.Yin}, imply that the D-brane charge can be interpreted in terms of K-theory. Consider a space-time manifold ${\cal{S}}$ and a D-brane wrapped in a $p+1$ dimensional submanifold ${\cal{W}}$ of it, so $f\,:\,{\cal{W}}\rightarrow {\cal{S}}$ denotes this embedding, and let ${\cal{N}}$ denote the normal bundle to ${\cal{W}}$. We can write the anomalous coupling as

\be
I_{\cal{W}}=\int_{\cal{W}}c\wedge Y\left( {\cal{F}},g\right),
\ee

\noindent where $c$ is the pullback of the total RR-field, $c=c^{(i)}+c^{(i+2)}+\dots$ with $i$ even for type IIB and odd for type IIA, and $Y$ is a function of the restriction of the space-time metric $g$ to the world-volume of the D-brane and ${\cal{F}}=F-B$, where $F$ is the $U(N)$ field strenght of the Chan-Paton vector field and $B$ is the restriction of the Neveu-Schwarz (NS) 2-form $B$ on the world-volume of the D-brane.

The subtleties come from the study of the function $Y$. In \cite{Gree.Har.Moo} it was found to be

\be
Y({\cal{F}},g)={\mbox{ch}}\left( E\right)f^*\left(\sqrt{\hat A\left({\cal{TS}}\right)}\right),
\ee

\noindent where ${\mbox{ch}}\left( E\right)$ is the Chern character of the Chan-Paton bundle, $E$, on the D-brane world-volume

\be
{\mbox{ch(E)}}=tr_N\exp\left(\frac{{\cal{F}}}{2\pi}\right).
\ee

\noindent ${\cal{TS}}$ denotes the tangent bundle to space-time ${\cal{S}}$, $f^*$ the pullback and $\hat A$ is the Dirac genus, defined as

\be
\hat A=\prod_{i=1}^n\frac{x^i/2}{\sinh x^i/2},
\ee

\noindent where $x_j=1+\Omega_j/2\pi$ and $\Omega_j$ is the field strenght of the connection on the ${\cal{TS}}$ bundle. 

However, when the normal bundle to the brane has non-trivial topology, $Y$ is modified, \cite{Min.Moo}, in order to include this effects

\be
\label{acc}
Y({\cal{F}},g)\longrightarrow Y({\cal{F}},g)e^{\frac{d}{2}}\frac{\hat A\left({\cal{TW}}\right)}{f^*\hat A\left({\cal{TS}}\right)},
\ee

\noindent where $d\in H^{2}\left({\cal{W},\ZZ}\right)$ is minus the first Chern class of the normal bundle, $c_1({\cal{N}})$, and its reduction the second Stiefel-Whitney class, $w_2({\cal{N}})$, therefore, its reduction mod 2 defines a Spin$^c$ structure on the world-volume ${\cal{W}}$.\footnote{This can be seen in terms of the Whitney sum rule, which implies that for an oriented and spin space-time manifold and an oriented world-volume, $w_2({\cal{N}})=w_2({\cal{W}})$.}

The equations of motion for the action including the couplin (\ref{acc}) allows us to write down an expression for the RR-charge, as an element of $H^*({\cal{S}})$, of the D-brane wrapping a supersymmetric cycle ${\cal{W}}$ as

\be
\label{dbc}
Q(E)={\mbox{ch}}\left( f_!E\right)\sqrt{\hat A\left({\cal{TS}}\right)},
\ee

\noindent where $f_!$ denotes the pullback for bundles.

Equation (\ref{dbc}) has a very nice interpretation in K-theory since it can be seen as a modified Chern character, giving a ring homomorphism

\be
{\mbox{ch}}\,:\,K^*(X,{\mathbb{Q}})\longrightarrow H^*\left( X,{\mathbb{Q}}\right),
\ee

\noindent that is is an isometry with respect to the usual pairing both in cohomology and in K-theory, the former being

\be
H^*\left( X,{\mathbb{Q}}\right)\otimes H^*\left( X,{\mathbb{Q}}\right)\longrightarrow {\mathbb{Q}},
\ee

\noindent given by

\be
\langle\omega_1,\omega_2\rangle_{DR}=\int_X\omega_1\wedge\omega_2,
\ee

\noindent while the pairing in K-theory is given in terms of the index theorem for the Dirac operator as

\be
\langle Q(E),Q(E')\rangle={\mbox{index}}\left( E\otimes E'\right).
\ee

The argument we have just given presents a formal evidence that D-brane charges should be classified in K-theory. However, this not the only one. Another argument is that Bott periodicity\footnote{see section \ref{ss22}} is reflected on the spectrum of type IIA, type IIB and type I string theories. Another crucial point is the fact that ordinary cohomology is not able to explain the shifted quantization rules for RR-fields (see for example \cite{Shift}) nor the precise multiplicative laws \cite{Wit.1}.

Moreover, K-theory is a natural framework to study the tachyon instabilities of Sen in what is known as the Sen-Witten construction. 

However, although very powerful, the study of processes of tachyon condensation may be considered as somehow speculative, since we have to turn to certain states that do not exist in the spectrum of our theory. This impels us to consider two more different points of view supporting this construction, the cancellation of global anomalies in the string world-sheet with D-brane boundary conditions \cite{Fre.Wit}, the Freed-Witten anomaly, and the topology of gauge fixings \cite{Gom} in terms of the 't Hooft's abelian projection \cite{THo.1}.

For vanishing B-fields, the Freed-Witten anomaly (see section \ref{ss23}) states that if the spinors cannot be defined globally, i.e. the second Stiefel-Whitney class $w_2({\cal{W}})$ does not vanish, the Dirac operator cannot be defined as a number and the D-brane world-volume must posses a Spin$^c$ structure in order to have a well defined string partition function. This is the same structure that appears in the anomalous couplings and leads again to an interpretation in K-theory, in terms of the Atiyah-Bott-Shapiro construction \cite{Ati.Bot.Sha}.

On the other hand, since K-theory classifies fibre bundles, one of its deepest implications is that there is a field theory underlying string theory which, in turn, states that the supergravity solutions should be put in correspondence with solitonic solutions of this field theory.

In type IIA we can interpret this gauge theory as the Yang-Mills theory defined on the world-volume of the system of the unstable D9-filling branes. This field theory is, in general, non-abelian, because of the enhancement of symmetry, and, in the simplest case, we obtain an abelian gauge theory, i.e. a single D-brane. However, this presents a subtlety and it is that even when the system does not incorporate lower dimensional charges, we have gauge structure in the closed string sector.

A way to gain some intuition on the physical meaning of this gauge theory is the 't Hooft's abelian gauge fixing. This construction shows how solitons arise as singularities in the context of a local gauge fixing, so it is tempting to conjecture that D-branes could be understood as the singularities arising from the gauge fixing in this underlying gauge theory \cite{Gom}.

The main focus of this paper is to review the role played by the B-field. Therefore, once the previous structure is presented, we will see how the presence of this NS 2-form modifies arguments leading to the twisted K-theory \cite{Ros, Don.Kar}, sometimes known as K-theory with local coefficients.

However, including the B-field forces us to be more carefuly since we find different algebraic structures depending on whether or not the B-field is flat (torsion). If the B-field is pure torsion, we can see that the theory on the D-brane world-volume is described in terms of certain algebras that are isomorphic to the set of matrices $M_m\left(\CC\right)$, \cite{Kap} , which can be interpreted as the algebraic statement that says that the B-field turns the world-volume gauge theory into a non-commutative one, \cite{Sei.Wit}.

However, when the B-field is non-torsion (for example in the presence of NS5-brane charges), we must consider a principal bundle over a separable Hilbert space and the previous algebra is replaced by the C$^*$-algebra of compact operators over this space \cite{Bou.Mat}. The most remarkable property of this bundle is its uniqueness which, when we restrict to the torsion subgroup, is lost and we obtain a non-unique locally trival bundle, depending on the choice of different but Morita equivalent algebras. 

Here we find another question which concerns the number of initial D9-branes needed to have a well defined construction. Suprinsingly, it seems that this number is indeed infinite, which is difficult to interpret physically. 

However, the mathematical set up predicts it very naturally. Let us look for the moment to ordinary $K^{-1}$ theory. In this case, the characteristic classes of the K-group are computed from maps to the infinite unitary group, \cite{Ati, Kar}. In twisted K-theory one adds to this structure the one carried by the H-field, which only changes $PU(\HS)$-equivariance, \cite{Ros, Bou.Mat}.

A proper interpretation of this state with infinite D9-filling branes seems to go over the problem of the $\ZZ_2$ symmetry that appears in the process of tachyon condensation \cite{Wit.4}. And, more interesting, can be related to the problem of cancellation of global anomalies in terms of K-homology \cite{Sza, Gom.Man.3} since it classifies both, cycles and gauge fields, and has, therefore, room enough to cover the structure of D-branes and allows us to define them as Fredholm modules (see for example, \cite{Mat.Sin}). Moreover, K-homology seems the natural way to understand the role played by the eleven dimensional $E_8$ bundles, \cite{Per.1}

In the last section of this review we will explore some connections between eleven dimensions and K-theory in the context of the D6-brane. As we will see, K-theory interprets the D6-brane as a non-abelian monopole. Our goal in this section will be the identification of the electric degrees of freedom as coming from the eleven dimensional 3-form, \cite{Gom.Man.2, Gom.Man.1}.

%%%%%%%%%%%%%%%%%%%%%%%%%%%%%%%%%%%%%%%%%%%%%%%%%%%%%%%%%%%%%%%%%%%%%%%%%%%%%%%%%%%%%%%%%%%%%%%%%%%%%%%%%%%%%%%%%%%%%%
%%%%%%%%%%%%%%%%%%%%%%%%%%%%%%%%%%%%%%%%%%%%%%%%%%%%%%%%%%%%%%%%%%%%%%%%%%%%%%%%%%%%%%%%%%%%%%%%%%%%%%%%%%%%%%%%%%%%%%
\section{K-theory in string theory}\label{s2}

In this section we will see how K-theory appears in the context of type II string theory giving rise to the whole spectrum of the theory. We will not review K-theory in a purely mathematical context, refering the reader to the extended literature on this subject (an incomplete list of references is \cite{Ati, Kar, Ros}, see also \cite{Ols.Sza}), but we will present it directly in the different contexts trying to give a clear exposition of the crucial concepts needed.

The organization is as follows. Firstly we will see how the processes of tachyon condensation leads to K-theory, then we will explain how the conditions for the cancellation of global anomalies in the string partition function leads to the same topological obstructions and so to the interpretation in terms of K-theory. Then, assuming that K-theory is the right theory for the classification of RR-charges, we will explore the physical meaning of the D-branes as singularities appearing in the gauge fixing.

%%%%%%%%%%%%%%%%%%%%%%%%%%%%%%%%%%%%%%%%%%%%%%%%%%%%%%%%%%%%%%%%%%%%%
%%%%%%%%%%%%%%%%%%%%%%%%%%%%%%%%%%%%%%%%%%%%%%%%%%%%%%%%%%%%%%%%%%%%%
\subsection{RR-charges in Type IIB}\label{ss21}

Although this paper will be mainly concerned with type IIA string theory and the predictions that the classification of RR-charges in terms of K-theory makes, it seems that the appearance of K-theory is more natural in type IIB, the reason being the existence in this theory of an RR 10-form to be associated with a D9-filling-brane. 

A known fact is that, due to the $SL(2,{\ZZ})$ symmetry of the equations of motion, type IIB string theory contains in its spectrum $(p,q)$ 9-branes carrying NS-NS 9-brane charge $p$ and RR 9-brane charge $q$ \cite{Hul}. Let us, then, take a system of N $D9$-branes and N' $\overline{D9}$-branes wrapped on a spacetime manifold $X$. The stack of $N$ $D9$-branes carry a $U(N)$ Chan-Paton bundle and the $N'$ $\overline{D9}$-branes carry a $U(N')$ Chan-Paton bundle, so the whole system possesses a $U(N)\times U(N')$ symmetry.

In this system we can have $(p$-$p)$, $(\bar p$-$\bar p)$ and $(p$-$\bar p)$ strings. The strings with its both ends in the same kind of brane have the usual GSO projection, while the $(p$-$\bar p)$ strings have the opposite one, which implies that the lowest mode in the NS-sector is a tachyon transforming in the $(N,\bar N')$ of $U(N)\times U(N')$ \cite{Gre, Per.2}.

The precise language representing this configuration is that of Quillen's superconnection \cite{Qui}, i.e. we represent the connection of this system of branes as

\be
A=\pmatrix{\nabla^+ & T \cr \bar T & \nabla^-},
\ee

\noindent where the $\pm$ signs stand for branes or anti-branes. This superconnection represents the lowest fields surviving the GSO projections. The $\nabla^{\pm}$ are connections coming from the open string sector with both ends on the branes (anti-branes) and have the usual GSO projection, while $T$ is the tachyon field, which comes from the sector with one end in each kind of brane and, therefore, have the opposite GSO connection.

The effect of this open string tachyon is to turn the system unstable and, therefore, it must decay into a stable state. This condensation process can lead either to an elementary state, i.e. a state whose charge does not differ from that of the closed string vacuum \cite{Sen.1}, or to a stable D-brane state. This last state carries a charge that is preserved under the addition of elementary charges, which is implemented in terms of creation or anihilation of elementary $(D9$-$\overline{D9})$ pairs \cite{Wit.1}.

It is worth to stress that what is said for $(D9$-$\overline{D9})$ pairs, can be said for $(Dp$-$\overline{Dp})$, for $p$ odd. However, it is useful to consider D9-branes as a starting point since then no configuration of lower dimensionality is lost, i.e. starting, say, from $(D7$-$\overline{D7})$ lying in the codimension $89$ plane, we lose all the posible D-brane configurations in this plane. Anyway, this is not a trivial statement, and the question under what conditions we can really say that any brane comes from this bound state will be studied in \ref{ss23} in the global context.

What we have said allows us to establish the following equivalence relation

\be
(E,F)\sim (E\oplus H,F\oplus H),
\ee

\noindent where $E$ and $F$ denote the Chan-Paton bundles of the initial system and $H$ denotes the Chan-Paton bundle of the elementary pair. However, this equivalence relation defines precisely the K-theory group of $X$: 

\be
\label{ktdrk}
K(X)=\tilde K(X)\oplus\ZZ,
\ee

\noindent where $\tilde K(X)$ is the reduced K-group defined as follows. Let us fix a basepoint in $X$ and take the inclusion $i\,:\, pt\hookrightarrow X$ and its induced map in K-theory $i^*\,:\, K(X)\rightarrow K(pt)=\ZZ$, then we can define $\tilde K(X)={\mbox{ker}}\,i^*$. The $\ZZ$ in (\ref{ktdrk}) represents the difference between the number of $D$ and $\overline D$-branes. However the cancellation of tadpole anomalies implies that the number of $D$-branes and $\overline D$-branes must be the same. All this indicates that D-brane charges in type IIB are classified by $\tilde K(X)$.

If the condensation process leads to a stable lower dimensional D-brane wrapping a submanifold $Y$ in spacetime, we can interpret the tachyon field as a Higgs type excitation. The dynamics of this condensation will then be such that the tachyon rolls down to the minimum of its potential, $T=T_0$, and breaks the gauge group down to its diagonal subgroup:

\be
U(N)\times U(N)\longrightarrow U(N).
\ee

\noindent The stable values of $T_0$ correspond to the vacuum manifold

\be
{\cal{V}}_{IIB}(N)=\frac{U(N)\times U(N)}{U(N)},
\ee

\noindent which is topologically equivalent to $U(N)$. There is, however, an unsolved puzzle, and it is the fate of this $U(N)$ in the case of an elementary pair, since we expected to end in the closed string vacuum which does not contain, in principle gauge groups \cite{Sre}.

The interesting point is that the stable soliton solutions will be classified in terms of maps from the compactly supported space-time manifold into the vacuum manifold, such that the charge of the soliton will be the winding of this map

\be
\pi_{m}\left({\cal{V}}_{IIB}(N)\right)=\ZZ,
\ee

\noindent and we are left with a D-brane wrapping a submanifold $Y$ of $X$. According to Bott's perdiodicity, the nontrivial windings are related to configurations with $m=2k$. In order to find out the relation between this $2k+1$ and the initial number of branes, we can consider what has been called the ``stepwise construction'' \cite{Wit.1}.

Following the initial construction of Sen, we look for a codimension two soliton solution in the core of the unstable system $(Dp$-$\overline{Dp})$, with gauge group $U(1)\times U(1)$. To give rise to a new system consisting in a pair $(D(p-2)$-$\overline{D(p-2)})$ we should start with a system of four branes, and this produces a codimension four object. This implies that we can recover a $Dp$-brane from a system of $2^{k-1}$ pairs of (p+2k)-branes and antibranes. 

We can then write the sequence

\be
\pi_{2k}\left( {\cal{V}}_{IIB}\left( 2^{k-1}\right)\right)=\pi_{2k-1}\left( U(2^{k-1})\right)=\ZZ.
\ee

Now we have to construct a generator of this map, i.e. the tachyon field. In order to do this, we follow the Atiyah-Bott-Shapiro construction \cite{Ati.Bot.Sha}. In the codimension $2k$ space we have a $SO(2k)$ group of rotations which can be lifted to give two spinor bundles: ${\cal{S}}_\pm$, of dimension $2^{k-1}$, which define a K-theory class via

\be
[{\cal{S}}_+]-[{\cal{S}}_-].
\ee

The tachyon vortex will be given by

\be
\label{tfftdb}
T(\vec x)=\vec\Gamma\cdot\vec x,
\ee

\noindent which has winding number 1, is a generator of $\pi_{2k-1}\left( U(2^{k-1})\right)$ and is a map between the two spin structures $T:\;{\cal{S}}_+\longrightarrow{\cal{S}}_-$. Indeed, (\ref{tfftdb}) is a generator of the compactly supported relative group $K\left( B^{2k},S^{2k-1}\right)$, where

\be
K\left( B^{2k},S^{2k-1}\right)=K\left( B^{2k}/S^{2k-1},pt\right)=\tilde K\left( S^{2k}\right).
\ee

%%%%%%%%%%%%%%%%%%%%%%%%%%%%%%%%%%%%%%%%%%%%%%%%%%%%%%%%%%%%%%%%%%%%%
%%%%%%%%%%%%%%%%%%%%%%%%%%%%%%%%%%%%%%%%%%%%%%%%%%%%%%%%%%%%%%%%%%%%%
\subsection{RR-charges in Type IIA}\label{ss22}

In order to describe the relevant K-groups for type IIA string theory, let us remind the existence of the so-called higher K-groups \cite{Kar}, defined to be

\be
K^{-n}(X)=K(\Sigma^nX),
\ee

\noindent where $\Sigma^nX\equiv S^n\wedge X$. In addition we have Bott's periodicity, which states

\be
\label{pdb}
K^{-n}(X)=K^{-n+2}(X).
\ee

It is tempting to say that type IIA string theory will be described in terms of equivalence classes of $K^{-1}$. Indeed, as $X=S^n$, then $\Sigma S^n=S^1\wedge S^n\simeq S^{n+1}$ and following the relation between K(X) and homotopy, we have

\bea
K^{-1}\left( S^{2n}\right) & = & 0, \\
K^{-1}\left( S^{2n+1}\right) & = & \ZZ.
\eea

This implies that we can have solitonic objects associated with a codimension odd space. These correspond precisely to the D-branes in type IIA. The definition we have just made of $K^{-1}(X)$ is known as the ``M-theory definition'' since it requires an extension of $X$ by an $S^1$. In section \ref{sss251} we will consider another definition which does not make use of this extension and will be refered as the ``stringy definition''.  

For computational purposes it is useful to set the relation with homotopy

\be
K^{-1}(X)=\left[ X,U(\infty)\right],
\ee

\noindent where $[A,B]$ denotes the classes of homotopy of maps $A\rightarrow B$ and $U(\infty)=\cup_{k=1}^\infty U(k)$ is the infinite unitary group, then it can be shown that

\bea
K^{-1}(S^n)=\pi_{n-1}\left( Gr(k,2k;\CC)\right),\qquad k>n,
\eea

\noindent where $Gr(k,2k;\CC)=\frac{U(2k)}{U(k)\times U(k)}$ are the Grassmannians, which represent a finite-dimensional approach to the universal classifying space $BU$. In view of this, we can say that this space of Grassmannians is the vaccum manifold for type IIA string theory. Let us see how to check this.

For our purposes it will be useful to introduce an unstable D9-brane \cite{Hor}. This may seem strange since there is no RR-charge associated to this object, and we could repeat the process as in type IIB although in this case we should start from a $(D8$-$\overline{D8})$ system. However, as we have already mentioned, this procedure would result in losing some configurations. 

The boundary state of this D9-brane is

\be
|D9\rangle=|D9,+\rangle_{NS}+|D9,-\rangle_{NS},
\ee

\noindent and, since the GSO projection acts as

\be
(-1)^{F_{L,R}}|D9,\pm>_R=|D9,\mp>_R,
\ee

\noindent on RR states, there is no GSO invariant combination of these states 

\be
P_{GSO}|D9\rangle_R=\left( 1-(-1)^{F_L}\right)\left( 1+(-1)^{F_R}\right)|D9\rangle_R=0.
\ee

\noindent This, in turn, implies that there is no condition on the cancellation of RR tadpole anomalies so we can have any number of 9-branes. In particular nothing prevents us from having an infinite number of them and this is the basis of some subtle questions concerning whether or not this infinite has a real physical meaning, although different arguments point to it. 

The condensation of the tachyon field depends on the explicit form of the tachyon potential, which is not known. However, from certain computations in open string field theory we can conclude that it is an even function of $T$ and that the minima can be written as $\pm T_0$. This implies that, after the condensation, the vacuum manifold is

\be
\label{vmftta}
{\cal{V}}_{IIA}(2N)=\frac{U(2N)}{U(N)\times U(N)}.
\ee

We can now look for the number of D9-branes needed to recover a codimension $(2k+1)$ D-brane. This number can be obtained again in two ways, the stepwise construction or in terms of the spinor bundle structure on the normal space to the brane world-volume. Both methods give the result $2N=2^k$. Then, the stable tachyon vortices are characterized by classes in

\be
\label{ghtda}
K^{-1}(S^{n+1})=\pi_n\left({\cal{V}}_{IIA}(2^k)\right)=\left\{\matrix{\ZZ &, & n=2k, \cr 0 &,& n=2k+1.}\right.
\ee

The precise form of the tachyon field will be determined, again, by the condition of being a generator of these homotopy groups. In this case, the system of $2^k$ D9-branes supports a $U(2^k)$ gauge theory. This group defines the Chan-Paton bundle and is taken to be the spinor bundle (which is now irreducible) of the group of rotations in the transverse space to the world-volume of the Dp-brane, $SO(2k+1)$. In this way the tachyon field will be a map $T\,:\,{\cal{S}}\rightarrow{\cal{S}}$, i.e. an automorphism of the Chan-Paton bundle on the D9-branes. A field satisfying all the conditions above is

\bea
T(x)=\Gamma_m x^m,\qquad 1\leq m\leq 2k+1.
\eea

This tachyon, although similar in form to (\ref{tfftdb}), has a different meaning, since it is now a generator of $K^{-1}\left( B^{2k+1},S^{2k}\right)$.

As the tachyon field corresponds to an automorphism of the spinor bundle, it is a field transforming in the adjoint representation of $U(2^k)$ and, therefore, carries no topology. Then a map like those taken place in (\ref{ghtda}) has only meaning if the number of D9-branes is indeed taken to be infinity. We can now recall, \cite{Kar}, that the $K{-1}(X)$ group can be defined in terms of such an inductive limit. Therefore, the infinite number of D9-branes has a perfect mathematical meaning. 

However, the definition of the vacuum manifold seems to imply that we only need a finite version of the $K^{-1}$-group, and we may wonder to what extent this is true. In fact, in order to properly define the D8-brane charge, we need this infinite number of initial unstable D9-branes, since this charge should be associated to a single homotopy group, which is

\be
\pi_0\left(O(1)\right)=\pi_0\left(\left\{\pm T_0\right\}\right)=\ZZ_2.
\ee

We inmediately see that this class does not represent a charge. The solution for this problem is taking the full $K^{-1}\left( S^1\right)=\ZZ$ and not the finite version of (\ref{ghtda}). Therefore, the group of $D8$-brane charges has an infinite number of filling branes \cite{Hor}.

The most important example for us will be the codimension three case, since this represents a D6-brane. Now we begin with a stack of two unstable $D9$-branes. The vacuum manifold corresponds to $U(2)$ gauge theory broken down to a $U(1)\times U(1)$ by the process of condensation

\be
{\cal{V}}=\frac{U(2)}{U(1)\times U(1)},
\ee

\noindent where the tachyon vortex takes the form 

\be
\label{tfctc}
T=x^i\sigma_i,
\ee

\noindent being the $\s_i$ the Pauli matrices. In this case, the relevant homotopy groups sequence is

\be
\label{sghct}
\pi_3(U(2))=\pi_2(U(2)/U(1)\times U(1))=\pi_1(U(1))=\ZZ.
\ee

We can now impose the finite energy conditions for this soliton, see \ref{ss61}, which tie it to the non-trivial $U(2)$ gauge field

\be
\label{sktds}
A_i(x)=\frac{1}{|x|^2}\left( 1-\frac{|x|}{\sinh|x|}\right)\Gamma_{ij}x^j.
\ee

The homotopy sequence and the gauge field imply that the D6 brane is a `t Hooft-Polyakov monopole for a $U(2)$ gauge theory \cite{Hor, Gom.Man.2}.

In the two previous examples, we see a main problem with the K-theoretic classification of RR-charges, and it is the appearance of a $\ZZ_2$ symmetry, either due to the graded distribution of the eigenvalues of the tachyon field or to the structure of the homotopy group. In the D8-brane case, this is solved by taking the whole group. For the D6-brane this also seems to be the solution, \cite{Wit.4}. However this solution is rather subtle, as we will see in section \ref{ss45}, and does not provide us with a complete solution.

%%%%%%%%%%%%%%%%%%%%%%%%%%%%%%%%%%%%%%%%%%%%%%%%%%%%%%%%%%%%%%%%%%%%%
%%%%%%%%%%%%%%%%%%%%%%%%%%%%%%%%%%%%%%%%%%%%%%%%%%%%%%%%%%%%%%%%%%%%%
\subsection{The global construction}\label{ss23}

All the treatment we have made in the previous sections relies in the assumption that we can define the spinor bundles globally, i.e. that the Chan-Paton bundles over the Dp-branes can be trivially extended over the whole space-time manifold $X$. However, in general this is not the situation. This section is devoted to review a construction which allows a proper definition of the K-groups in such situation \cite{Wit.1, Ati.Bot.Sha, Ols.Sza}.

Let us discuss the situation in the type IIB context. The previous construction can be described in the following terms. Let us take a space-time manifold $X$ and a complex line bundle $\LL$ over it. Now suppose we are given a submanifold $Y$ of $X$ which will represent a Dp-brane. The tachyon field will be a section of $\LL$ vanishing at $Y$ and of constant length otherwise. The Dp-brane charge of $Y$ can be seen to arise if we consider a $D$-$\overline D(p+2)$ system such that the D-brane has a $U(1)$ connection on $\LL$ and the $U(1)$ over the $\overline D$-brane is trivial. It is clear that the tachyon is a section of this $U(1)\times U(1)$.

Now suppose we want to include in $Y$ lower dimensional charges (this is the same of asking under what conditions all D-branes in the spectrum can arise from a system of $N$ $(D9$-$\overline{D9})$-branes) then, over $Y$ we will have another line bundle $\MM$. If this $\MM$ can be extended over the whole $X$ the Dp-brane can be described in the same way as before but now taking on the $D(p+2)$-brane the bundle $\LL\otimes\MM$ and on the $\overline D(p+2)$-brane $\MM$.

In the general case, however, $\MM$ does not extend over $X$ and we have to use a construction due to Atiyah-Bott-Shapiro by taking a tubular neighborhood $Y'$ of $Y$. Let $(E,F)$ two vector bundles over $Y$ that define a class in $K(Y)$ and by means of the inclusion map we can take this class to $K(\overline Y)$ on the closure of $Y'$. Then the tachyon field, being $T\,:\,F\rightarrow E$, will be an isomorphism when restricted to the the boundary $Y^*$. If $F$ can be extended from the boundary to $X'=X-Y'$ then it can be defined over the whole $Y$ and by means of the isomorphism with $E$ on $Y^*$ it can also be extended over the whole $X$. However, if $F$ cannot be extended, we can take another vector bundle $H$ such that $F\oplus H$ is trivial over $Y$ and over $\overline Y$. Now we can define the K-theory group $K(Y)$ in terms of the replacement

\be
(E,F;T)\rightarrow (E\oplus H,F\oplus H;T\oplus 1).
\ee

\noindent In the case of type IIB the bundles $(E,F)$ will simply be the spinor bundles ${\cal{S}}_{\pm}$. However, another obstruction appears if we cannot define these spinor bundles.

In order to understand this obstruction, let us consider the following exact sequence

\be
0\stackrel{\a}{\longrightarrow} \ZZ \stackrel{\g}{\longrightarrow} 2\ZZ \stackrel{\b}{\longrightarrow} \ZZ_2 \stackrel{\d}{\longrightarrow} 0,
\ee

\noindent where exactness means that the kernel of an application is the image of the previous one, so this sequence reads as follows. The first map takes the zero into the zero in the integers implying that ${\mbox{ker}}\, \g=\left\{0\right\}$ so $\g$ is injective. On the other hand $\b$ being onto, is an epimorphism, since ${\mbox{ker}}\, \d=\left\{\ZZ_2\right\}$ and this is the image of $2\ZZ$ under the mod 2 reduction $\b$.

This sequence induces a cohomology long exact sequence

\be
\dots H^2\left( X,2\ZZ\right)\stackrel{\b^*}{\longrightarrow} H^2\left( X,\ZZ_2\right)\stackrel{\pd^*}{\longrightarrow}H^3\left( X,\ZZ\right)\stackrel{\g^*}{\longrightarrow} H^3\left( X,2\ZZ\right)\stackrel{\b^*}{\longrightarrow}\dots,
\ee

\noindent where $\pd^*$ is the connecting homomorphism. Now let us take characteristic classes $c_1\in H^2(X,\ZZ)$, $w_2\in H^2(X,\ZZ_2)$ and $W_3\in H^3(X,\ZZ)$, which are respectively the first Chern class, the second Stiefel-Whitney class and the third integer Stiefel-Whitney class. We use again the exactness of this sequence, which now says that if $W_3=\pd^*\left( w_2\right)$. Then $w_2$ lifts to $c_1$ mod 2 iff $W_3=0$, which is precisely the condition for a Spin$^c$ structure (see e.g. \cite{Osc}). 

This is easily seen by recalling that the obstruction to lift $SO(n)$ to $Spin(n)$ is this $w_2$

\be
0\longrightarrow \ZZ_2 \longrightarrow Spin(n) \longrightarrow SO(n) \longrightarrow 0.
\ee

\noindent However, if this class is the reduction mod 2 of an integer cohomology class we can construct the Spin$^c$ bundle where

\be
0\longrightarrow U(1) \longrightarrow Spin^c(n) \longrightarrow SO(n) \longrightarrow 0,
\ee

\noindent and the Spin$^c$ group will be a twisted bundle ${\cal{S}}\otimes{\cal{L}}$ that can be written as

\be
\label{scm}
Spin^c(n)=Spin(n)\times U(1)/\ZZ_2.
\ee

In the language of transitions functions of the bundle, the fact that $w_2\neq 0$ implies that the cocycle relation satisfied is

\be
g_{ij}g_{jk}g_{ki}=\phi_{ijk},
\ee

\noindent where $g_{ij}\in SO(n)$ and $\phi_{ijk}\,:\, U_i\cap U_j\cap U_k\rightarrow \ZZ_2$ is the 2-cochain defining $w_2$. Then the twisting by ${\cal{L}}$ is such that it cancels this 2-cochain and defines a good bundle. This is possible if ${\cal{L}}=K^{1/2}$ has transition functions staisfying $f_{ij}=\pm h_{ij}^{1/2}$ with $h_{ij}h_{jk}h_{ki}=\phi_{ijk}$, then the transition funcions $g_{ij}h_{ij}$ close to one in the cocycle. 

The relation with K-theory can now be established by saying that the tachyon is a map between these Spin$^c$ bundles. For type IIB the tachyon will be a map $T\,:\,{\cal{S}}_+\otimes K^{1/2}\rightarrow {\cal{S}}_-\otimes K^{1/2}$ and as for type IIA we don not have this graded spin bundles, the tachyon will be again an automorphism $T\,:\,{\cal{S}}\otimes K^{1/2}\rightarrow {\cal{S}}\otimes K^{1/2}$. 

This global construction implies that a brane can wrap a supersymmetric cycle ${\cal{W}}$ only if $W_3({\cal{W}})=0$, although we are not imposing any conditions for more than one D-brane, where it could be possible anyway.

%%%%%%%%%%%%%%%%%%%%%%%%%%%%%%%%%%%%%%%%%%%%%%%%%%%%%%%%%%%%%%%%%%%%%
%%%%%%%%%%%%%%%%%%%%%%%%%%%%%%%%%%%%%%%%%%%%%%%%%%%%%%%%%%%%%%%%%%%%%
\subsection{The Freed-Witten anomaly}\label{ss24}

In this section we will adopt what could be considered as a more conservative point of view, that of cancellation of anomalies \cite{Fre.Wit}. The conditions we will find will be exactly the same we have already found in section (\ref{ss23}) concerning the obstructions to have a globally defined construction. 

The Freed-Witten anomaly arises in the string theory path integral from open Riemann surfaces that end on the D-brane. In this section we will deal only with the case of vanishing Kalb-Ramond B-field, postponing the study of this case until section (\ref{ss33}) where we have already defined all the precise mathematical tools involving this field.

Once the fermions are integrated, the relevant factors in the string path integral are

\be
\label{spiwbf}
{\mbox{pfaff}}\left( D\right)\cdot\exp\left( i\oint_{\pd\Sigma}A\right),
\ee

\noindent where ${\mbox{pfaff}}\left( D\right)$ is the Pfaffian (the square root of the determinant) of the Dirac operator of the Rarita-Schwinger field, and $A$ is the Chan-Paton vector field in the $U(1)$ gauge theory on $Q$.

The study of index theorems for families and determinant line bundles (let us denote by ${\cal{L}}$ the determinant line bundle, which is a smooth complex line bundle) gives a geometric picture that states that ${\cal{L}}$ has a natural connection whose curvature represents the local anomaly and its holonomy the global one \cite{Wit.3, Bis.Fre, Fre}. 

This holonomy is found as follows. Let us take a loop $\g\,:\,S^1\rightarrow X$, then the $S^1$ parametrizes the family of Dirac operators obtained by pullback. If we couple the Dirac operator to a vector bundle $\pi\,:\, E\rightarrow X$. We endow the $S^1$ with a metric and require independence on it, which means that we rescale it by $g^{(S^1)}/\epsilon$ and see the behaviour in $\epsilon\rightarrow 0$. This procedure is an adiabatic limit. The holonomy is then taken to be

\be
{\mbox{hol}}\,{\mbox{Det}}\left( D^{\gamma^{-1}}(E)\right)={\mbox{a-lim}}\,e^{2\pi i\xi},
\ee

\noindent where $\xi=\frac{\eta_E+h_E}{2}$, being $h_E$ the dimension of the kernel of the Dirac operator and $\eta_E$ the eta invariant, defined by analytic continuation as the value of

\be
\eta_D\left[ s\right]=\sum_{\lambda\in{\mbox{spec}}\left( D\right)\backslash\left\{ 0\right\}}{\mbox{sgn}}\left(\lambda_i\right)|\lambda_i|^{-s},
\ee

\noindent at $s=0$ \cite{Ati.Pat.Sin}.

This determinant bundle has a natural square root, the Pffafian line bundle. The main result in \cite{Fre.Wit} is that the sign of the Pfaffian cannot be well defined as a number. 

In fact, let us take type II string theory on a spacetime $X$, and take a one parameter's family of world-sheets $\Sigma$ parametrized by a circle $C$, we can define the map $\phi\,:\,\Sigma\times C\rightarrow X$ and $\phi\left(\pd\Sigma\times C\right)\in Q$. When we go around the loop $C$, the holonomy of the Pfaffian is the sign factor

\be
\label{hofp}
\left(-1\right)^{\left(\pd\Sigma\times C,\phi^*\left( w_2(Q)\right)\right)},
\ee

\noindent where $w_2(Q)$ is the second Stiefel-Whitney class of $Q$. This power can be rewritten as

\be
\a=\int_{\pd\Sigma\times C}w_2(Q)=\left(\pd\Sigma\times C,w_2(Q)\right)=\left(\pd\Sigma\times C,w_2(\n)\right)=,
\ee

\noindent once we consider $\phi\,:\,\pd\Sigma\times C\rightarrow Q$ as an embedding, and where we have used the Whitney sum rule in the last equality, for $\n$ the normal bundle.

This clearly implies that if $w_2(Q)$ is non-zero, the Pfaffian is not well defined as a number, which implies that the second term in (\ref{spiwbf}) must have the same ambiguity in order to have a well-defined path integral. Tracing back to the previous considerations on Spin$^c$ manifolds, we see that this is the same condition, i.e. if we cannot define spinors globally, the bundle over the D-brane world-volume admits a Spin$^c$ structure.

The compensating ambiguity in sign can be obatined from the spin connection in terms of the following product

\be
\label{scagf}
{\mbox{Tr}}P\exp\left(\oint_{\pd\Sigma}\omega\right)\cdot\exp\left( i\oint_{\pd\Sigma}A\right),
\ee

\noindent where $\omega$ is the spin-connection, which has $SO(N)$ as group structure and whose holonomy is defined up to a sign due to the double cover of $Spin(N)$. The product considered in (\ref{scagf}) is then well defined.

Equation (\ref{scagf}) represents spinors of charge 1 with respect to $A$, and so they are not sections of the spin bundle associated but of the twisting of it by the hermitian line bundle of which $A$ is a section. Then the connection on the $Spin^c$ bundle over $Q$ can be written as $\omega+A$.

%%%%%%%%%%%%%%%%%%%%%%%%%%%%%%%%%%%%%%%%%%%%%%%%%%%%%%%%%%%%%%%%%%%%%
%%%%%%%%%%%%%%%%%%%%%%%%%%%%%%%%%%%%%%%%%%%%%%%%%%%%%%%%%%%%%%%%%%%%%
\subsection{Another point of view: the Abelian Projection}\label{ss25}

The usual perturbation theory is blind to the compactness of the gauge group and, in this way, it does not see certain relevant variables related with non-trivial topological effects, such as instantons or monopoles. In \cite{THo.1} a ghost free unitary gauge was proposed in order to isolate these variables such that one could have a deeper knowledge of the fields involved in the confinement problem. The key point is that certain singularities, seen as Gribov ambiguities in the gauge fixing, have the physical meaning of being precisely those variables.

The way to proceed is to fix the gauge as locally as possible. To perform this gauge fixing, let us take an extra field $X$ transforming in the adjoint representation of the gauge group, then it is a $N\times N$ hermitian matrix and, generaly, traceless. The most important feature of this field is its transformation law, not the field itself:

\be
X\rightarrow X=\Omega X\Omega^{-1},
\ee

\noindent in such a way that it does not involve either derivatives of $\Omega$ nor $\Omega$ evaluated at different points.

As the eigeinvalues of $X$ are gauge invariant, we will choose the gauge such that $X$ is diagonal. To make this we can introduce some Lagrange multipliers to cancel the non-diagonal terms:

\be
{\cal L}_{gauge}=\sum_{i<j}\a_{ij}X^{ij}.
\ee

\noindent This leaves $X$ as

\be
X=\pmatrix{\lambda_1&&0\cr
                     &\ddots&\cr
                     0&&\lambda_N},
\ee

\noindent where the eigenvalues will be given certain order presciption. For example, if $X$ lives in the Lie algebra of $SU(N)$, we can write  $\l_1\geq\l_2\geq...\geq\l_N$.

However, this does not fix completely the gauge, due to the fact that $\Omega$ is defined up to multiplicative factors

\bea
d={\mbox{diag}}\left(\exp(i\a_1),\exp(i\a_2),...,\exp(i\a_N)\right),\quad\sum_i\a_i=0,
\eea

\noindent where $(d)$ is the larger Cartan, or abelian, subgroup: $(d)=U(1)^{N-1}\in SU(N)$, and forms a residual local gauge group: $U(1)^{N}/U(1)=U(1)^{N-1}$. The transformed gauge field of $X$ will be $X$ again only if it commutes with $\Omega$, which, in turn, is only possible if it is also diagonal: $(d)$.

This residual gauge symmetry can be fixed as in QED by means, for example, of the Lorentz gauge:

\be
{\cal L}^{gauge,abe}=\sum_{i=1}^{N-1}\b_i\pd_\m A^\m_{ii},
\ee

\noindent so the total gauge fixing Lagrangian takes the form

\be
{\cal L}_{gauge}=\sum_{i<j}\a_{ij}X^{ij}+\sum_{i=1}^{N-1}\b_i\pd_\m A^\m_{ii},
\ee

\noindent which gives us the dimension of the group of transformations in terms of the number of Lagrange multipliers: $2^{\frac{N(N-1)}{2}}+N-1$.

Now we can see that we have all the features of an abelian gauge theory with $N-1$ fold multiplicity, i.e. with gauge group $U(1)^{N-1}$. The residual gauge transformations act in the field as

\bea
(A_\m)_{ii} & \rightarrow & (A_\m)_{ii}-\frac{1}{g}\pd_\m\Lambda_{ii},\\
(A_\m)_{ij} & \rightarrow & \exp(i(\Lambda_i-\Lambda_j))(A_\m)_{ij}.
\eea

As the diagonal gauge fields transforms as $N$ gauge potentials with the extra condition $\sum_i(A_\m)_{ii}=0$, we will call them photons. All the other fields transform as $N(N-1)$ charged vector fields. We will name this non-diagonal photons electrically charged gluons. 

The existence of these fields is important because in a theory like QCD we have electrically charged quarks and gluons and a mechanism for confinement will take into account both fields. Moreover, these gluons are massive, due to the fact that they are not protected by gauge invariance because we have eliminated the non-diagonal symmetry

\bea
(D_\m X)_{ij} & = & \pd_\m X_{ij}+ig(\l_i-\l_j)(A_\m)_{ij},\\
{\mbox{Tr}}(D_\m X)^2 & \rightarrow & (\pd_\m X)^2+g^2(\l_i-\l_j)^2(A^{ij}_\m)^2.
\eea

Fortunatelly, the theory is not exactly $U(1)^{N-1}$, something has survived of its non-abelian nature, which resides in the singularities of the gauge fixing, which appear whenever $\l_i=\l_j$, because in this case the composite eigenvectors of $\Omega$ are not well defined. Then $\Omega$ has a line of directional singularities, which can be interpreted as the worldline of a magnetic monopole. For example, in the $SU(2)$ case, the conditions that the hermitian matrix

\be
X=a_0{\cal{I}}+a_i\s^i,
\ee

\noindent has two coinciding eigenvalues is

\be
a_i=0,
\ee

\noindent i.e., three conditions. This example serves us as worm up exercise for the more general case.

Let us take $X$ as the Higgs field for a grand unification theory with the symmetry spontaneously broken $SU(N)\rightarrow U(1)^{N-1}$. Near the singularity, it will take the form

\be
\label{xns}
X=\pmatrix{\ddots &    &    & 0 \cr
                  & \l & 0  &   \cr
                  &  0 & \l &   \cr
                0 &    &    & \ddots} + \sum_{k=1}^3a_k(x)\pmatrix{\ddots &      & \cdots \cr
                                                                          &\s_k  &        \cr
                                                                   \cdots &      &     \ddots},
\ee

\noindent where we have made use of a parametrization of the $SU(2)$ subset in terms of the Pauli matrices. The first term on the r.h.s of (\ref{xns}) is gauge invariant, while the second one will vanish, as in the previous example, when we approximate some subspace where

\bea
x\rightarrow x_0: a_k(x)\rightarrow 0\quad k=1,2,3.
\eea

Generically these three conditions eliminate three planes meeting at $x_0$, which fixes, in three dimensions, a point, while in four dimensions fixes a (world) line. 

Now we can make $a_k(x)=(x-x_0)_k$ and, at $x_0$, the residual symmetry is enhanced from $U(1)^{N-1}$ to $U(1)^{N-3}\times U(2)$, which is non-abelian, and these degrees of freedom are magnetic monopoles with respect to this symmetry.

This is not difficult to see, because the parametrization used near the singular point allows us to define a projector from the field

\be
X=\sum_{i=1}^3x_i\s^i,
\ee

\noindent at infinity, as

\be
\Pi_{\pm}=\frac{1}{2}\left(1\pm X\right),
\ee

\noindent which, in turn, allows us to decompose the trivial bundle $S^2\times C^2$ in two different line bundles $E_{\pm}$, taking into account the projection over $C^2$, and the magnetic monopoles are defined by the associated principal bundles, with charges $g_i=(0,...0,,1-1,0,...0)$.

These magnetic monopoles will acquire mass since there is nothing to prevent them from getting it, so electromagnetism provides the only long range fields, the $N-1$ $U(1)$ photons.

%%%%%%%%%%%%%%%%%%%%%%%%%%%%%%%%%%%%%%%%%%%%%%%%%%%%%%%%%%%%%%%%%%%
%%%%%%%%%%%%%%%%%%%%%%%%%%%%%%%%%%%%%%%%%%%%%%%%%%%%%%%%%%%%%%%%%%%
\subsubsection{K-theory and the Abelian projection}\label{sss251}

In sections \ref{ss21}-\ref{ss23} we have described D-branes in String Theory as certain solitons arising from the condensation of the tachyon field, interpreted as some kind of Higgs excitation. However, in the previous description of the singularities in the gauge fixing, we have found the equation (\ref{xns}), which allowed us to identify the extra degrees of freedom in the non-abelian theory as magnetic monopoles. This equation resembles (\ref{tfctc}), which allowed us to interpret the D6-brane as a 't Hooft-Polyakov monopole. It seems tempting to interpret D-branes as singularities of the gauge fixing and lift this interpretation to K-theory \cite{Gom}.

This interpretation is possible, as we already mentioned in the introduction, since we will interpret the K-theoretical classification of RR-charges as indicative of some field theory degrees of freedom underlying string theory. In type I string theory one could say that this gauge theory is given in terms of the $SO(32)$ gauge symmetry, however it is far from clear what should it be in type II.

Whatever it is, it is clear that this gauge theory, as formulated is related to the open string sector via Chan-Paton bundles and open string tachyon condensation, and we do not know what modifications arise when we finally understand how to implement the closed string sector.

In the next chapters, however, we will deal with a first step towards this direction, and it is the inclusion of the NS  2-form field. As we will see the modifications it brings are conceptually far from trivial although the general picture remains, i.e. we can still consider D-branes as solitons in a dynamical process of tachyon condensation.

The way we make contact with 't Hooft's abelian gauge fixing is in terms of the stringy definition of $K^{-1}(X)$, \cite{Kar, Hor}. Let us begin defining a pair: $\left( I^n,\a\right)$ consisting in the set $I^n$ of trivial vector bundles on a compact manifold $X$ and the automorphism, $\a$, of $I^n$. 

We will say that such a pair is elementary when the automorphism $\a$ is homotopic to the identity within the automorphisms of $I^n$.With this definition we can establish an equivalence relation between two pairs $\left( I^n,\a\right)$ and $\left( I^m,\b\right)$ as

\be
\label{ertdskt}
\left( I^n\oplus I^p,\a\oplus\g_1\right)\sim\left( I^n\oplus I^q,\a\oplus\g_2\right),
\ee

\noindent where $\left( I^{p,q},\g_{1,2}\right)$ are two elementary pairs. This means that equivalence classes, i.e. the elements of $K^{-1}$ are going to be related to the homotopy classes of the automorphism $\a$.

The set of these equivalence classes on $X$ forms a group, the inverse being $\left( I^n,\a^{-1}\right)$, so that $\left(I^n\oplus I^n,\a\oplus\a^{-1}\right)$ is an elementary pair, which is precisely the higher K-theory group $K^{-1}(X)$. We will not try to probe rigurously here the equivalence between the two different definitions given for $K^{-1}$, which can be found, for example, in \cite{Kar} Theorem II.4.8. 

In the context of string theory, the role of the vector bundle is played by the Chan-Paton bundle carried by the system of $N$ unstable D9-filling branes, and the automorphism $\a$ will be \cite{Hor}

\be
\label{ukt}
{\cal{U}}=-e^{i\pi T},
\ee

\noindent where $T$ is the adjoint $U(N)$ tachyon on the filling branes world-volume. The elementary pairs, then, will correspond to elementary configurations and so can be created from and anhiliated to the vacuum.

The important point here is that (\ref{ukt}) defines a map

\be
{\cal{U}}\,:\,S^{2k+1}\rightarrow U(2^k),
\ee

\noindent and, consequently, an element of $\pi_{2k+1}\left( U(2^k)\right)$, which defines the bound state construction of D-branes in type IIA as stated in section \ref{ss22}.

The unitary ghost free gauge fixing described by the Abelian projection defines in a natural way an automorphism of the corresponding gauge bundle. In fact, once we fix the non-abelian part of the gauge by diagonalizing the field $X$, we can define the automorphism as

\be
\a(x)=e^{iX(x)},
\ee

\noindent and the K-charge is defined by the homotopy class of this $\a(x)$ \cite{Gom}.

However, if we are going to take this analogy seriously, there is a crucial ingredient to be included, and it is the stability of the topological charge in the abelian projection with respect to the equivalence relation (\ref{ertdskt}). This can be seen as passing from the gauge group $U(N)$ to $U(M)$, with $M>N$. However, in the abelian projection, the meaning of the topological charge associated to the $U(N)$ subgroup is independent of $M$, and so is stable under creation of elementary $D9$ filling branes.

%%%%%%%%%%%%%%%%%%%%%%%%%%%%%%%%%%%%%%%%%%%%%%%%%%%%%%%%%%%%%%%%%%%%%%%%%%%%%%%%%%%%%
%%%%%%%%%%%%%%%%%%%%%%%%%%%%%%%%%%%%%%%%%%%%%%%%%%%%%%%%%%%%%%%%%%%%%%%%%%%%%%%%%%%%%
%%%%%%%%%%%%%%%%%%%%%%%%%%%%%%%%%%%%%%%%%%%%%%%%%%%%%%%%%%%%%%%%%%%%%%%%%%%%%%%%%%%%%
\section{Non-trivial B-fields}\label{s3}

For the purposes of this paper, and in order to see how to include the B-field in the previous discussion, it is interesting to gain some geometrical and topological intuition on it. This can be done by considering the following double spectral sequence

\begin{center}
\begin{tabular}{c c | c c c c c c c}  
 $\Omega^3$ & $H$ & $dB_\a$ & & & & & & \\
& & $\uparrow$ & & & & & & \\ 
 $\Omega^2$ &  & $B_\a$  &  $\rightarrow$  & $\d B_\a=d\psi_{\a\b}$ & & & & \\ 
& & & & $\uparrow$ & & & &\\ 
 $\Omega^1$ &  & & & $\psi_{\a\b}$ & $\rightarrow $ &  $\d\psi_{\a\b}=dg_{\a\b\g}$ & &    \\ 
& & & & & & $\uparrow$  & &\\ 
 $\Omega^0$ &  & & & & &  $g_{\a\b\g}$  &  $\rightarrow$   &  $\d g_{\a\b\g}$ \\ 
% & & & & & & & &\\ 
\hline 
& &  & & & & & & $m_{\a\b\g\d}$ \\
& & $U_\a$  & &  $U_{\a\b}$  & & $U_{\a\b\g}$ & & $U_{\a\b\g\d}$ \\
\end{tabular} 
\end{center}

From this diagram we can read the gauge structure involving the B-field. It reads as follows. The outter columm sets the de Rham cohomology, i.e. for differential forms, while the outter row sets the \v Cech cohomology, i.e. for cocycles. Obviously, in this outter regions the Poincar\'e lemma is not satisfied, and so not every global closed differential form is exact and exactly the same for the \v Cech cohomology on the rows \cite{Orl}. Then we see that on the columns acts the exterior derivative $d$ and in the rows acts the coboundary operator $\d$. The nomenclature $\Omega^i$ denotes the space of differential $i$-forms and $U_{\a\b\dots}$ is the intersection $U_\a\cap U_\b\cap\dots$ of open covers.

This double spectral sequence says that, in order to describe properly the B-field, we need the triple $\left(B_\a,\psi_{\a\b},g_{\a\b\g}\right)$. Following \cite{Hit}, we can translate the previous diagram into explicit expressions.

Let us consider a manifold $X$ and an open covering on it. Defining a B-field, $B_\a$, we have

\be
B_\b-B_\a=d\psi_{\a\b},
\ee 

\noindent where the $\psi_{\a\b}$ are 1-forms defined on the double intersection. In the triple intersection they will satisfy

\be
\d\left\{\psi_{\a\b}\right\}=\left\{\psi_{\a\b}+\psi_{\b\g}+\psi_{\g\a}\right\}=\left\{ dg_{\a\b\g}\right\}.
\ee

Let us now take a set of $U(1)$-valued 0-forms defined on the triple intersection

\be
\label{gcr}
f_{\a\b\g}:U_\a\cap U_\b\cap U_\g\longrightarrow S^1,
\ee

\noindent in such a way that

\be
f_{\a\b\g}=\exp\left( ig_{\a\b\g}\right),
\ee

\noindent which satisfy $f_{\a\b\g}=f^{-1}_{\b\a\g}=f^{-1}_{\a\g\b}=f^{-1}_{\g\b\a}$ on $U_{\a\b\g}$. Using these 0-forms we set

\be
\psi_{\a\b}+\psi_{\b\g}+\psi_{\g\a}=(-i)f^{-1}_{\a\b\g}df_{\a\b\g},
\ee

\noindent which allows us to interpret the 1-forms $\psi_{\a\b}$ as connections of a line bundle defined on each double intersection. On the other hand, these $f$`s also satisfy the cocycle relation

\be
f_{\b\g\d}f^{-1}_{\a\g\d}f_{\a\b\d}f^{-1}_{\a\b\g}=1,
\ee

\noindent which, in terms of $g_{\a\b\g}$, defines a \v Cech cochain

\be
\d\left\{ g_{\a\b\g}\right\}=im_{\a\b\g\d}.
\ee

In analogy with the Dirac monopole, where quantum consistency implies that the cochain takes values in $\mathbb{Z}$, we can argue that

\be
\left\{\frac{m_{\a\b\g\d}}{2\pi}\right\}\in{\mathbb{Z}},
\ee

\noindent so $\left\{\frac{m_{\a\b\g\d}}{2\pi}\right\}$ defines a \v Cech cochain in the integers and $m$ is a \v Cech cocyle which represents a class $[H]$ in $H^3\left( X,{\mathbb{Z}}\right)$. In this sense we can interpret the B-field as a connection on a gerbe.

The $f_{\a\b\g}$ take values in $Cont_X\left(U(1)\right)$, the sheave of continuous functions on $X$ with values in $U(1)$ (remember that this sheave assigns to every open set in $X$ the abelian group of continuous functions from each $U_\a$ to $U(1)$ with pointwise multiplication as group operation). Then the exact sequence of sheaves

\be
0\longrightarrow{\mathbb{Z}}\longrightarrow Cont_X\left({\mathbb{R}}\right)\longrightarrow Cont_X\left(U(1)\right)\longrightarrow 0,
\ee

\noindent induces the long exact sequence in cohomology

\bea
\nonumber \dots\longrightarrow H^2\left( X,Cont_X\left( {\mathbb{R}}\right)\right) & \longrightarrow & H^2\left( X,Cont_X\left( U(1)\right)\right)\longrightarrow \\
\nonumber \\
& \longrightarrow & H^3\left( X,{\mathbb{Z}}\right)\longrightarrow H^3\left( X,Cont_X\left( {\mathbb{R}}\right)\right)\longrightarrow\dots,
\eea

\noindent which, due to the fact that $Cont_X\left( {\mathbb{R}}\right)$ is a fine sheave with vanishing cohomology groups, leads to 

\be
\dots\longrightarrow 0\longrightarrow H^2\left( X,Cont_X\left( U(1)\right)\right)\longrightarrow H^3\left( X,{\mathbb{Z}}\right)\longrightarrow 0\longrightarrow\dots.
\ee

\noindent It is this isomorphism the one which maps the cohomology class of $f_{\a\b\g}$ to the class $[H]$ in $H^3(X,\ZZ)$.

Now we have a complete description of the gerbe asocciated to a generic B-field in terms of the triple $(B_\a,\psi_{\a\b},g_{\a\b\g})$. We can define the trivializations of the gerbe in terms of functions defined on the double cover

\be
f_{\a\b}=f^{-1}_{\a\b}:U_\a\cap U_\b\longrightarrow S^1,
\ee

\noindent such that $f_{\a\b\g}=f_{\a\b}f_{\b\g}f_{\g\a}$ and the difference between two such trivializations is a flat line bundle. In this way, we can define a flat line bundle on each double intersection which will have a connection, $\nabla_{\a\b}$, such that

\be
\d B_\a=d\nabla_{\a\b},
\ee

%%%%%%%%%%%%%%%%%%%%%%%%%%%%%%%%%%%%%%%%%%%%%%%%%%%%%%%%%%%
%%%%%%%%%%%%%%%%%%%%%%%%%%%%%%%%%%%%%%%%%%%%%%%%%%%%%%%%%%%
\subsection{Torsion B-fields}\label{ss31}

Supose now that we are given a submanifold of spacetime, $Q$, such that the restriction of the B-field to it is trivial in the de Rham cohomology. In other words, the B-field in that subspace is flat and defines a pure torsion characteristic class. Let us now see how the double spectral sequence is modified in this case

\begin{center}
\begin{tabular}{c c | c c c c c c c}  
 $\Omega^3$ &  & $0$ & & & & & & \\
& & $\uparrow$ & & & & & & \\ 
 $\Omega^2$ &  & $B_\a$  &  $\rightarrow$  & $\d B_\a=d\psi_{\a\b}$ & & & & \\ 
& & $\uparrow$ & & $\uparrow$ & & & &\\ 
 $\Omega^1$ &  & $A_\a$ & $\rightarrow$ & $\psi_{\a\b}$ & $\rightarrow $ &  $\d\psi_{\a\b}=dg_{\a\b\g}$ & &    \\ 
& & & & $\uparrow$ & & $\uparrow$  & &\\ 
 $\Omega^0$ & & & & $\rho_{\a\b}$  & $\rightarrow$ &  $g_{\a\b\g}+c_{\a\b\g}$  &  $\rightarrow$   &  $\d (g_{\a\b\g}+c_{\a\b\g})$ \\ 
% & & & & & & & &\\ 
\hline 
& &  & & & & & & $\tilde m_{\a\b\g\d}$ \\
& & $U_\a$  & &  $U_{\a\b}$  & & $U_{\a\b\g}$ & & $U_{\a\b\g\d}$ \\
\end{tabular} 
\end{center} 

This structure defines a flat gerbe, where the B-field is the field strength of certain 1-form field, i.e. $B_\a=dA_\a$ in $U_\a$. In this case the equations deduced in section \ref{ss21} have to be modified. Firstly, the variation of the B-field now takes into account this A-field as

\be
\d B_\a=B_\b-B_\a=d\psi_{\a\b}=d\left( A_\b-A_\a\right),
\ee

\noindent which implies

\be
\label{gtftg}
\psi_{\a\b}+A_\a-A_\b=d\rho_{\a\b}.
\ee

On the other hand, it is not difficult to see that

\be
d\left(\d\rho_{\a\b}-g_{\a\b\g}\right)=0,
\ee

\noindent and, in turn

\be
\d\rho_{\a\b}=g_{\a\b\g}+c_{\a\b\g},
\ee

\noindent where the $c_{\a\b\g}\in 2\pi\RR/\ZZ$ is a \v Cech cochain that defines a class in $H^2(Q,U(1))$ in $Q$, where this $U(1)$ represents now a group and not a sheave. This class is what we can associate with the holonomy of the connection. This defines a flat gerbe with a holonomy class in $H^2\left( Q,\RR/\ZZ\right)$. Using the exact sequence of groups

\be
0\longrightarrow \ZZ\longrightarrow \RR\longrightarrow U(1)\longrightarrow 0,
\ee

\noindent the induced long exact cohomological sequence

\be
\dots\longrightarrow H^2\left( Q,\RR\right)\longrightarrow H^2\left( Q,U(1)\right)\longrightarrow H^3\left( Q,\ZZ\right)\longrightarrow H^3\left( Q,\RR\right)\longrightarrow\dots,
\ee

\noindent leads to the isomorphism

\be
H^2\left( Q,U(1)\right)\cong H^3\left( Q,\ZZ\right).
\ee

\noindent This is a Bockstein homomorphism, $\b$, and can be written as

\bea
\label{ac}
\left.\matrix{\xi & \in & H^2\left( Q,U(1)\right) \cr [H] & \in & H^3\left( Q,\ZZ\right)} \right\}\longrightarrow\b(\xi)=[H],
\eea

\noindent where it should be understood that $[H]\in Tors\left( H^3(Q,\ZZ)\right)$. In section \ref{ss43} we will see how this Bockstein is related to the cancellation of global anomalies.

%%%%%%%%%%%%%%%%%%%%%%%%%%%%%%%%%%%%%%%%%%%%%%%%%%%%%%%%%%%%%%%%%%%
%%%%%%%%%%%%%%%%%%%%%%%%%%%%%%%%%%%%%%%%%%%%%%%%%%%%%%%%%%%%%%%%%%%
\subsection{The $D=10$ KK-monopole}\label{ss32}

Our purpose now will be the application of the ideas explained in the previous section to the case of the KK-monopole, which is known to have an electric charge associated to the B-field \cite{Sen.3} (see section \ref{sss631}), and we would like to see how the electric and magnetic charges are related in its geometry. 

The gerbe associated to this object is flat, with $B=dA$, where

\be
A=\frac{Cr}{4m(r+4m)}\left[ dx^5+4m(1-\cos\theta)d\phi\right].
\ee

In this case, one can consider two 1-forms which are identified with the gauge potentials associated to the electric and magnetic charges. They come from the components of the metric and the B-field along the compact direction of the Taub-NUT space:

\bea
A_\m^1 \equiv g_{\m 5} & = & 4m(1-\cos\theta)d\phi, \\
A_\m^2 \equiv B_{\m 5} & = & \frac{\tilde C}{(r+4m)^2}dr.
\eea

We can consider that the source for the B-field is a string winding along the compact direction which, when it follows an unwinding trayectory, will produce the electric charge of the KK-monopole \cite{Greg.Har.Moo}. With this in mind, we can decompose the B-field as

\be
B_{\m\n}={\cal{B}}_{\m\n}+\frac{1}{2}\left( A_\m^1A_\n^2-A_\n^1A_\m^2\right).
\ee

For our present purposes, we will cover the $S^3$ at infinity only by two patches, as given by the transition needed to avoid the singularities of the metric $x^4\rightarrow x^4\pm 8m\phi$. However, we should say that, as the intersection is now non-contractible, we cannot reach the \v Cech cocycle.

The gauge transformations for the fields will be

\bea
\tilde A^1 & \rightarrow & A^1-8m\phi, \\
\tilde B   & \rightarrow & B+4md\phi\wedge A^2,
\eea

\noindent which produce

\be
\label{elkk}
\psi_{\a\b}=8m\phi\frac{\tilde C}{(r+4m)^2}dr,
\ee

\noindent and

\be
\rho_{\a\b}=-2\tilde C\left(\frac{r\phi}{r+4m}\right).
\ee

Notice that the 1-form (\ref{elkk}) is defined on the intersection of the two patches which is homotopy equivalent to $S^2$ and, in consequence, is a $U(1)$-connection over the $S^2$.

%%%%%%%%%%%%%%%%%%%%%%%%%%%%%%%%%%%%%%%%%%%%%%%%%%%%
%%%%%%%%%%%%%%%%%%%%%%%%%%%%%%%%%%%%%%%%%%%%%%%%%%%%
\subsection{The D6-brane}\label{ss33}

The previous example for the ten dimensional KK-monopole can be easily adapted to the eleven dimensional one, i.e. to the D6-brane (see section \ref{ss64}). In this case, a crucial point is that the B-field we will consider is, when the dimensional reduction is taken into account, partially dictated by the geometry of the space-time and partially by the eleven dimensional 3-form, $C^{(3)}$. Moreover, we can see that the D6-brane is coupled to the string ending on its world-volume, which implies, as we will see in section \ref{ss64} that the B-field on the brane is pure torsion, \cite{Gom.Man.1}.

Again, the gerbe structure can be computed in a very simple way. We will follow the notation in \cite{Gom.Man.1}, although not it normalizations, i.e. $A$ and $B$ denote the fields on the world-volume of the D6-brane and $V$ and $C$ the fields on its transverse space. Therefore, in the presence of the KK-monopole, the eleven dimensional 3-form $C^{(3)}$ contains $C_{\m\n i}=C_{\m\n}A_i$ and $C_{\m ij}=V_\m B_{ij}$ as interesting factors, where the latin indices take values in the transverse space and greek ones in the world-volume. Now we impose an harmonicity condition on the zero modes in the transverse space, which implies \cite{Ima, Gom.Man.2}

\be
\label{tsce}
C\propto dV,
\ee

\noindent where now $V$ can be written in terms of a pair of functions $f_1(r)$ and $f_2(r)$, whose explicit form is not relevant for the purposes of this section (we will come back to this in section \ref{sss652}) as

\be
V=f_1(r)dr+\frac{f_2(r)}{2}\left(\cos\theta\pm 1\right)d\phi+\frac{f_2(r)}{2m}dx^4.
\ee

Now we can perform the transformation $x^4\rightarrow x^4\pm 8m\phi$ and, chosing the minus sign, we find that

\be
\d V=-2\a f_2(r)d\phi,
\ee

\noindent where we have introduced a constant $c$ which should be fixed by physical considerations. Computing $d\left(\d V\right)$ we find that it can also be written as the exterior derivative of another 1-form, namely

\be
d\left(\d V\right)=-2\a\frac{\pd f_2(r)}{\pd r}dr\wedge d\phi=d\left( 2\a\frac{\pd f_2(r)}{\pd r}\phi dr\right)\equiv d\Lambda.
\ee

\noindent It is very easy to check that $\d V$ and $\Lambda$ differ by an exact form $d\rho$, where $\rho$ is

\be
\label{ftcb}
\rho_{\a\b}=-2\a f_2(r)\phi.
\ee

This would give us a complete description of the gerbe in the transverse space to the KK-monopole. However, we can gain more information from the 4-form, $G_{\m\n ij}$. We can write

\be
\d G_{\m\n ij}=\d C_{\m\n}(dA)_{ij}+C_{\m\n}\d(dA)_{ij}+\d(dV)_{\m\n}B_{ij}+(dV)_{\m\n}\d B_{ij},
\ee

\noindent and, using now (\ref{tsce}), this can be put as

\be
\d G_{\m\n ij}=C_{\m\n}\left(\d dA+\d B\right)_{ij}.
\ee

\noindent This equation sets the usual relation between the variations of the Chan-Paton and the Kalb-Ramond fields on the world-volume of the brane, namely

\bea
B & \longrightarrow & B + d\Lambda, \\
A & \longrightarrow & A - \Lambda + d\rho,
\eea

\noindent which defines the flat gerbe structure on the world-volume of the D6-brane.

%%%%%%%%%%%%%%%%%%%%%%%%%%%%%%%%%%%%%%%%%%%%%%%%%%%%%%%%%%%%%%%%%%%%%%%%%%%%%%%%%%%%%%%%%%%%%%%%%%%%%%%%%%%%%%%%%%%%%%%%%%%%%%
%%%%%%%%%%%%%%%%%%%%%%%%%%%%%%%%%%%%%%%%%%%%%%%%%%%%%%%%%%%%%%%%%%%%%%%%%%%%%%%%%%%%%%%%%%%%%%%%%%%%%%%%%%%%%%%%%%%%%%%%%%%%%%
\section{K-theory with B-fields}\label{s4}

In this section we will explain how to construct the twisted K-groups of a given manifold. As stated in \cite{Wit.1, Kap, Bou.Mat} these groups are relevant when a B-field is included and depends on the gerbe structure defined on the manifold. 

As we have already done in the previous section, we have to consider two different situations, a non-torsion or a torsion B-field. In the first case the bundle structure that is defined is unique and given in terms of an infinite Hilbert space, while in the second case the bundle is locally trivial, non-unique and defined in terms of a finite dimensional algebra isomorphic to $M_m(\CC)$.

%%%%%%%%%%%%%%%%%%%%%%%%%%%%%%%%%%%%%%%%%%%%%%%%%%%%
%%%%%%%%%%%%%%%%%%%%%%%%%%%%%%%%%%%%%%%%%%%%%%%%%%%%
\subsection{Non-torsion twisted K-groups}\label{ss42}

In this section we deal with non-torsion B-fields. As we have seen in the previous section, the torsion B-field in the world-volume of a D-brane appears as a consequence of the harmonicity imposed on the zero modes in the transverse space. For a D9-brane there is no transverse space enough to define any 2-form. Therefore, we can consider that the B-field living in its world-volume is non-torsion.

In \cite{Bou.Mat} it was proposed that the Rosenberg's definition of twisted K-theory, \cite{Ros}

\be
\label{tkg}
K^j\left(X,[H]\right)=K_j\left(\Gamma_0\left(X,{\cal{A}}_{[H]}\right)\right),
\ee

\noindent where $\Gamma_0\left(X,{\cal{A}}_{[H]}\right)$ is the algebra of sections vanishing at infinity of the continuous field ${\cal{A}}_{[H]}$ of elementary C$^*$-algebras over $X$, is the relevant one when we consider non-torsion B-fields. And its relation with homotopy is

\be
\label{tkgh}
\matrix{ K^0\left(X,[H]\right) & = & \left[ P_{[H]},U({\cal{Q}})\right]^{PU} \cr && \cr 
         K^{-1}\left(X,[H]\right) & = & \left[ P_{[H]},U\right]^{PU}  },
\ee

\noindent where $U({\cal{Q}})$ is the unitary group of the C$^*$-Calkin algebra ${\cal{Q}}={\cal{B}}({\cal{H}})/{\cal{K}}$, i.e. the quotient space between the algebras of bounded and compact operator on the Hilbert space ${\cal{H}}$, and $U$ is the group of unitary operators in the unitalization of ${\cal{K}}$

\be
U=\left\{ u\in U({\cal{H}})\big| u-1\in{\cal{K}}\right\}.
\ee

These equivalence classes can be computed following the Atiyah-J\"anich theorem, \cite{Ati, Jan}, which states that the space of Fredholm operators $\FF$ is the classifying space for the functor $K^0$ and from the Atiyah-Singer theory for skew-adjoint Fredholm operators, \cite{Ati.Sin}, which states that the classiying space for the functor $K^{-1}$ is $\FF_*$, see for example \cite{Boo.Ble} for an expository account. Now we proceed to define all the elements of these statements.

As we said in section \ref{s3}, the gerbe for a generic B-field has a characteristic class $[H]=H^3\left( X,\ZZ\right)$, which can be computed from

\be
H^3\left( X,\ZZ\right)=\left[ X,K(\ZZ,3)\right],
\ee

\noindent i.e. it is related to the homotopy classes from the manifold to the third Eilenberg-Mac Lane space. In \cite{Gom.Man.3} we will investigate this relation in order to point the relation with $E_8$ bundles. Now, following the Dixmier-Douady theorem for continuous-trace algebras \cite{Dix.Dou}, we can search another model for this space. Let us consider an infinite dimensional separable Hilbert space $\HS$ and the group of unitary operators on it. The group $U(1)$ will consist on scalar multiples of the identity operator on $\HS$ with unit norm. This set can be written as $\RR/\ZZ$ and is a $K(\ZZ,1)$. 

We can look for the classifying space $BU(1)$. This will be $PU(\HS)=U(\HS)/U(1)$ since it is the base space of a locally trivial $U(1)$-bundle with contractible total space, so is itself a model for $K(\ZZ,2)$. Following these lines, we can say that $BPU(\HS)$ is a model for $K(\ZZ,3)$, and then view the characteristic class $[H]$ as the homotopy class of maps

\be
\phi\,:\, X\longrightarrow K(\ZZ,3)\simeq BPU(\HS),
\ee

\noindent or as defining a $PU(\HS)$-principal bundle

\be
PU(\HS)\longrightarrow P_{[H]}\stackrel{p}{\longrightarrow} X
\ee

From this principal bundle we can build its associated vector bundle as ${\cal{A}}_{[H]}=P_{[H]}\times_{PU(\HS)}\KK$, where $PU(\HS)$ acts on $\KK$ by $^*$-automorphisms, given by $Ad\,:\,T\longrightarrow gTg^{-1}$, i.e. in terms of the adjoint map, which is known to be a continuous homorphism of $U(\HS)$ onto ${\mbox{Aut}}(\KK)$ with kernel $U(1)$. This can be stated by saying that the group of autmorphisms of $A_{[H]}$ which fix the spectrum $X$ pointwise is given by:

\be
{\mbox{Aut}}_XA_{[H]}\simeq\left\{ PU(\HS){\mbox{-equivariant continuous maps }}P_{[H]}\longrightarrow PU_{Ad}\right\}.
\ee

The sections of this vector bundle that vanish at infinity form an algebra from whose spectrum we recover the base manifold $X$ which we will denote as $A_{[H]}=\Gamma_0\left( X,P_{[H]}\times_{PU(\HS)}\KK\right)$.

The interesting point of the Dixmier-Douady theorem is that if this algebra $A_{[H]}$ is stable, i.e. $A_{[H]}\simeq A_{[H]}\times\KK$. Then ${\cal{A}}_{[H]}$ is locally trivial with fibers $\simeq\KK$ and $A_{[H]}$ is determined, up to automorphisms fixing $X$ pointwise, by $[H]$ and any such class $[H]$ arises from a (unique) stable separable continuous-trace algebra $A_{[H]}$ over $X$. 

It is worth to mention that, since $\KK\otimes\KK\simeq\KK$, the isomorphism classes of these bundles form a group under the tensor product with inverse the conjugate bundle. This group is called the infinite Brauer group and denoted by $Br^\infty(X)$. We will give a local coordinate description of all these facts in the next section in the torsion B-fields context.

The Gelfand-Na\u\i mark theorem, \cite{Gel.Nai} states that $A_{[H]}$ is the algebra of the compact topological space $X$, which at the same time be reconstructed from the spectrum of the algebra by means of the the Gelfand transformation. This implies that we can compute the K-groups in terms of the space $X$ or in terms of $A_{[H]}$, which is precisely the definition made in (\ref{tkg}). However, for computational purposes, we would like to relate the classes of these groups with classes of homotopy maps. 

\paragraph{{\sl{type IIA \\[0.5cm]}}}

The relevant K-group for type IIA string theory is $K^{-1}$, therefore, let us remind that $K_1(A_{[H]})$ is defined as the group of path-components of

\be
\left\{ u\in U(A_{[H]}^+)\,\big|\,u-1\in A_{[H]}\right\},
\ee

\noindent which is isomorphic to the sections $X\rightarrow P_{[H]}\times_{PU(\HS)} U$, i.e. to the $PU(\HS)$-equivariant maps $P_{[H]}\rightarrow U$, so that we can compute the equivalence classes from (\ref{tkgh}).

Going back to the Atiyah-Singer theorem for skew-adjoint Fredholm operators, we can say that the space of self-adjoint operators is homotopycally equivalent to the set of unitary operators of the for $u=1+k$, where $k\in\KK$, \cite{Ati.Sin}. In this way we write

\be
\label{tkgfo}
K^{-1}\left( X,[H]\right)=\left[ P_{[H]},\FF_*\right]^{PU(\HS)},
\ee

\noindent being the mapping between these two sets of operators given by

\be
\label{uyfo}
g=-e^{i\pi F},
\ee

\noindent where $g-1\in\KK$ and $F\in\FF_*$.

Let us recall the definition of $K^{-1}(X)$ made in section \ref{sss251} in terms of pairs $(I^n,\a)$, and let us now define the $K^{-1}(X,[H])$ in the same way. In our case, we have to consider maps not from the base manifold but from the total space, so we can write \cite{Gom.Man.3}

\be
\Phi\,:\,P_{[H]}\stackrel{\pi}{\longrightarrow}X_{cpt}\stackrel{\a}{\longrightarrow}U,
\ee

\noindent then we take a point $x\in X_{cpt}$ and its image under $\pi^{-1}$ so we have

\be
\pi^{-1}(x)=\left. PU(\HS)\right|_x.
\ee

We can, on the other hand, think of the map $\Phi$ as the exponentiation of of the C$^*$-algebra of global sections, which are given by a collection of functions $f_\a$ of the form $f_\a\,:\, U_\a\longrightarrow \KK$ transforming in the adjoint, i.e. on double overlaps they satisfy $f_\a=Ad(g_{\a\b})f_\b$, with $g_{\a\b}\,:\, U_\a\cap U_\b\rightarrow U(\HS)$ and $Ad(g_{\a\b})\,:\, U_\a\cap U_\b\rightarrow Aut(\KK)=PU(\HS)$, as

\be
\label{pitotace}
\Phi=-e^{i\pi f}.
\ee

It is not difficult to see from (\ref{pitotace}) that the condition of $PU(\HS)$-equivariance implies that $PU(\HS)$ acts, effectively, as automorphims of $U_{cpt}$. Therefore, we can propose, at least locally, that the vacuum manifold for type IIA in the presence of a non-torsion B-field takes the form

\be
\label{vmwntbf}
{\cal{V}}_{IIA}=\frac{U_{cpt}}{PU(\HS)}
\ee

Details and modifications of this construction are beyond the scope of this review, and we refer the reader to \cite{Gom.Man.3}.

\paragraph{{\sl{type IIB \\[0.5cm]}}}

In this case we use the group $K_0(A_{[H]})$, which can be defined as group of the path components of the unitary groups of ${\cal{Q}}_\d=\Gamma\left( X,P_{[H]}\times_{PU(\HS)}{\cal{Q}}\right)$, \cite{Ros}. However, there is another definition that fits better in our understanding of the noncommutative tachyon field (see section \ref{ss44}), and it is as the Grothendieck group of Murray-Von Neumann equivalence classes of projections, $P_\a$, in $A_{[H]}\otimes\KK$ \cite{Bou.Mat}.

These projections in $\KK$ are a collection of functions $P_\a\,:\, U_\a\rightarrow \KK$ and define a finite range subspace, $V_{\a,x}$, in the Hilbert space $\HS$, such that on a double intersecction $U_\a\cap U_\b$ an element $v\in V_{\a,x}$ is identified with $g_{\a\b}(x)v\in V_{\a,x}$. We can now define a gauge bundle as the data $\left( U_\a,\left\{ V_{\a,x}\right\}_{x\in U_\a},g_{\a\b}\right)$. 

Now, we remember that the Murray-Von Neumann equivalence states that two such projectors, $P_\a$ and $Q_\a$, are equivalent if there exists another projector, $\Lambda_\a$ such that $P=\Lambda^*\Lambda$ and $Q=\Lambda^*\Lambda$. In terms of gauge bundles we can define

\be
K^0\left( X,[H]\right)=\left\{ \left[\left\{ U_\a,\left\{ V_{\a,x}\right\}_{x\in U_\a},g_{\a\b}\right\}\right]- \left[\left\{ U_\a,\left\{ W_{\a,x}\right\}_{x\in U_\a},g_{\a\b}\right\}\right]\right\},
\ee

\noindent such that $\left[\left\{ U_\a,\left\{ V_{\a,x}\right\}_{x\in U_\a},g_{\a\b}\right\}\right]$ and $\left[\left\{ U_\a,\left\{ W_{\a,x}\right\}_{x\in U_\a},g_{\a\b}\right\}\right]$ are gauge bundles over $X$ with the former being defined by $P_\a$ and the second by $Q_\a$ and such that $\Lambda_\a\,:\,V_{\a,x}\rightarrow W_{\a,x}$ is an isomorphism of these gauge bundles.

%%%%%%%%%%%%%%%%%%%%%%%%%%%%%%%%%%%%%%%%%%%%%%%%%%%%%%%%%%%%%%%%%%%%%%%%%%%%%%%%%%%%%%%%%%%%%%%%%%%%%%%
%%%%%%%%%%%%%%%%%%%%%%%%%%%%%%%%%%%%%%%%%%%%%%%%%%%%%%%%%%%%%%%%%%%%%%%%%%%%%%%%%%%%%%%%%%%%%%%%%%%%%%%
\subsection{Torsion twisted K-groups}\label{ss41}

Let us now come back to equation (\ref{gtftg}). This gauge transformation of $A$ does not define a gauge transformation for a connection due to the $\Lambda$ term. Thus in order to interpret it as a gauge connection, we must get rid of it. In doing this, we use the fact \cite{Hit} that the difference between two different trivializations of a gerbe is a flat line bundle, and take like in \cite{Kap} a new connection such that:

\be
\m\longrightarrow\m-\Lambda+dg,
\ee

\noindent with $g$ satisfying $g_{\a\b}g_{\b\g}g_{\g\a}=f_{\a\b\g}$ as in (\ref{gcr}). Then, we can define the new gauge connection as

\be
\tilde A=A-\m,
\ee

\noindent such that it satisfies the usual relation:

\be
\d\tilde A=id\log h_{\a\b},
\ee

\noindent where $h_{\a\b}=m_{\a\b}\a_{\a\b}^{-1}$ with $m_{\a\b}=e^{ig_{\a\b}}$ and $\a_{\a\b}=e^{i\rho_{\a\b}}$.

However, these new transition functions does not satisfy the usual cocycle relation, instead they obey

\be
\label{craa}
h_{\a\b}h_{\b\g}h_{\g\a}=\xi_{\a\b\g},
\ee

\noindent where we have imposed the condition that the holonomy of the gerbe on the brane, $Q$, is trivial, so that the coefficients $c_{\a\b\g}$ are a coboundary.

Now, to make contact with string solitons, we can consider a stack of $N$ filling branes (again a system in type IIB can be more natural), and $Q$ as its world-volume. Then the matrices $h_{\a\b}$ are seen as $U(N)$ valued. However, relation (\ref{craa}) says that the bundle on the brane world-volume is not a $U(N)$, since the cocycle relation for its transition functions does not close to one but to an element of the center of the group, namely to a $U(1)$ element, which is only inmaterial if we go to the adjoint representation, which has as group structure $U(N)/U(1)=SU(N)/\ZZ_N$.

Equation (\ref{craa}) is also the defining relation of an Azumaya algebra, which is a locally trivial algebra over $X$ with fibre isomorphic to the algebra $M_m\left(\CC\right)$ of matrices $m\times m$ over the complexes. This implies that the world-volume of these D-branes is described in terms of a module $\Gamma$ over an Azumaya {\bf A} defined by the matrices $h_{ij}$ such that its representative class is $\d{\mbox{\bf A}}$, \cite{Don.Kar}. Moreover, the set of all equivalence classes of an Azumaya algebra over $X$ is called the Brauer group of $X$, denoted by $Br(X)$ which is isomorphic to $Tors\left( H^3(X,\ZZ)\right)$ by a theorem of Serre \cite{Gro}, and we have the relation $\d{\mbox{\bf A}}=\b(\xi)=[H]$. The sections of the bundle we have obtained consist of functions $R_\a:\,{\cal{U_\a}}\rightarrow u(N)$, where $u(N)$ is the Lie algebra of $SU(N)/\ZZ_N$ and on double overlaps they satisfy

\be
\label{aotb}
R_\a=h_{\a\b}R_\b h_{\a\b}^{-1}=Adj(h_{\a\b})R_\b,
\ee

\noindent which represents an automorphism of the sections of the bundle equal to its structure group.

Now, noticing that direct sum of two twisted bundles as described above is again a twisted bundle, we can establish the equivalence relation $(E,F)\sim (E\oplus H,F\oplus H)$ which defines the equivalence class in $K_{[H]}(X)$, \cite{Don.Kar}. This is precisely the twisted K-theory used in \cite{Wit.1, Kap}. 

The K-groups for type IIA can be equally defined as in section \ref{ss42}, where now the automorphisms of the bundle are defined to act in the sections of this bundle as in (\ref{aotb}). 

The main difference with the construction of the previous section is that now the bundle is not unique, therefore given two twisted bundles of those described above, the corresponding algebras will be different but Morita equivalent. This equivalence in turn implies that their K-groups equal each other, i.e. $K\left({\bf{A}}\right)=K\left({\bf{A}}'\right)$.

%%%%%%%%%%%%%%%%%%%%%%%%%%%%%%%%%%%%%%%%%%%%%%%%%%%%%%%%%%%%%%%%%%%%%%%%%%%%%%%%%%%%%%%%%
%%%%%%%%%%%%%%%%%%%%%%%%%%%%%%%%%%%%%%%%%%%%%%%%%%%%%%%%%%%%%%%%%%%%%%%%%%%%%%%%%%%%%%%%%
\subsection{Anomaly cancellation with B-fields}\label{ss44}

Let us now go back to the analysis of section (\ref{ss24}), but including the Kalb-Ramond field. The bosonic part of the partition function will now contain the following factors

\be
\label{cotbfpf}
\exp\left( i\oint_{\pd\Sigma}A+i\int_\Sigma B\right),
\ee

\noindent in the presence of D-brane boundary conditions, we interpret (\ref{cotbfpf}) by saying that the A-field provides a trivialization for the B-field, which is viewed as a section of a line bundle, ${\cal{L}}_B$, over the space of Riemann surfaces that end on the D-brane, i.e. when restricted to the space of loops on the world-volume of the D-brane, $LQ$. The important point is that such a trivialization only exists if $H=dB$ is cohomologically trivial when restricted to the D-brane, and can be seen from the general results in section (\ref{ss31}) as the usual gauge transformation

\be
B\rightarrow B-d\Lambda,\qquad A\rightarrow A+\Lambda+d\rho.
\ee

If we now include the contribution of the fermions, the trivialization must be for the twisted Pffafian bundle ${\mbox{Pfaff}}\otimes{\cal{L}}_B$ when restricted to $LQ$. 

In this case, the cancellation of the global anomaly stems from the proper identification of the first Chern classes involved. On one hand, the integral of the characteristic class $\left. [H]\right|_Q$ over $\pd\Sigma$ yields the first Chern class ${\cal{L}}_B$

\be
c_1({\cal{L}}_B)=\oint_{\pd\Sigma}\left. [H]\right|_Q.
\ee

On the other hand, the class of the Pfaffian as a flat line bundle is the integral of $w_2(Q)$ over $\pd\Sigma$, see equation (\ref{hofp}), therefore, its first Chern class can be seen in terms of the Bockstein homomorphism as

\be
c_1({\mbox{Pffaf}})=\b\left(\oint_{\pd\Sigma}w_2\right)=\oint_{\pd\Sigma}W_3(Q).
\ee

In this way we obtain the condition for cancellation of global anomalies \cite{Fre.Wit, Wit.9}

\be
\label{cogawbf}
\left. [H]\right|_Q=W_3(Q).
\ee

In reference \cite{Kap} we can find also the non-abelian generalization, which includes the 't Hooft flux characteristic class $y\in H^2(Q,\ZZ_n)$, with $n$ the number of D-branes

\be
\left. [H]\right|_Q=W_3(Q)+\b\left( y\right).
\ee

Equation (\ref{cogawbf}) implies that a single D-brane can only wrap a supersymmetric cycle, $Q$, if this condition is satisfied. However, there is a further refinement due to possible instabilities carried by instantons \cite{Mal.Moo.Sei}. This refinement says that a brane wrapping $Q$ can nevertheless be unstable if for some $Q'\subset X^9$

\be
\mbox{PD}\left( Q\subset Q'\right)=W_3(Q')+\left.[H]\right|_{Q'}.
\ee

Obviously, a D-brane wrapped in $Q'$ is unstable. However, a brane wrapping a codimension 3 cycle $Q\subset Q'$ provides a magnetic source that cancells the anomaly. The physical picture, \cite{Mal.Moo.Sei} is that a brane wraps a spatial cycle $Q$ propagating in time and ends in a D-instanton wrapping $Q'$. The most interesting question in this case is that we gain a physical interpretation of twisted K-theory, which can be stated \cite{Moo} as ``cancellation of anomalies modulo instanton effects''. However, although very interesting, this picture is still incomplete, since it makes use of the Atiyah-Hirzebruch spectral sequence (AHSS) which is an algorithm that gives only an approximation to $K_{[H]}$. In a forthcoming paper \cite{Gom.Man.3} we propose that these conditions can be properly understood if one considers K-homology groups instead.

%%%%%%%%%%%%%%%%%%%%%%%%%%%%%%%%%%%%%%%%%%%%%%%%%%%%%%%%%%%%%%%
%%%%%%%%%%%%%%%%%%%%%%%%%%%%%%%%%%%%%%%%%%%%%%%%%%%%%%%%%%%%%%%
\subsection{Noncommutative tachyons}\label{ss43}

One of the more striking effects of the B-field on the theory in the D-branes world-volume is that it produces a non-commutative deformation (\cite{Sei.Wit} and references therein) in the sense of Connes \cite{Con}. The point here is that the B-field can be related to a length scale $\theta^{ij}$ defined from the commutator

\be
\left[ x^i,x^j\right]=\theta^{ij},
\ee

\noindent which depends with $B$ as

\be
\label{ddtcb}
\theta^{ij}=2\pi\a'\left(\frac{1}{g+2\pi\a'B}\right)^{ij}_A\stackrel{(g,\a')\rightarrow 0}{\longrightarrow}\left(\frac{1}{B}\right)^{ij},
\ee

\noindent where $(\,)_A$ denotes the antisymmetric part of the matrix and $i,j=1,\dots,r$, and $r$ the rank of $B$.

This length scale appears in many interesting problems, for example, in the resolution of singularities in the moduli space of instantons, the so-called small instanton singularities \cite{Cal.Har.Str}. Here the noncommutativity parameter sets a ``no-go'' scale preventing us from going to zero size. A resolution ot this type was suggested by Nakajima in reference \cite{naka} and corresponds to the replacement of the real moment map $\m_R=0$ in the ADHM construction, \cite{adhm}, by $\m_R=constant$. Nekrasov and Schwarz noticed in \cite{Nek.Sch} (see also \cite{Sei.Wit}) that this resolution of the small instanton singularity of the instanton moduli space exactly corresponds to the non-commutative deformation of the ADHM real moment map being the constant deformation the non-commutative parameter.

More interesting for our purposes is the monopole case. Now the moduli space is non-singular and presents the mathematical structure of a Hilbert scheme, $X^{[k]}$, with $X$ the moduli space $\mathbb{R}^3\times S^1$ of a $k=1$ monopole. The Hilbert scheme $X^{[k]}$ is a desingularization of ${\cal{S}}^k(X)$, the symmetric product of $X$ $k$ times, that would be the natural moduli space of $k$-monopoles interpreted as a set of $k$ different particles \cite{naka, Ati.Hit}.

In the non-commutative case, the real moment map $\m_R$ is modified to $\m_R=-\theta$ \cite{Gom.Man.2}. However, what is interesting is that, as we have already mentioned, we do not need this deformation of $\m_R$ to desingularize the moduli space. The physical reason for this is that even without $B$ field, the elementary constituents of a $k$-monopole are at short distances delocalized.

The noncommutative Yang-Mills theory is also non-local by virtue of this length scale. This is reflected in terms of the multicplicative structure which is given in terms of a Moyal or a Kontsevich \cite{Kon} deformation for constant or non-constant B-field respectively. We will consider only a Moyal deformation, given by the Moyal product, symbolized as usual by a ``$\star$''

\be
\left( A\star B\right)=\left. e^{\frac{\theta}{2}\left(\pd_z\pd_{\bar z'}-\pd_{z'}\pd_{z'}\right)}A(z,\bar z)B(z',\bar z')\right|_{z=z'},
\ee

Our first step in the construction of a K-theoretical classification with B-fields is considering this noncommutative theory for a scalar field, which will be taken to be the tachyon field \cite{Gop.Min.Str, Das.Muk.Raj, Har.Moo}. The main result in this context is that, in certain limit to be specified later, the tachyon field is a projector, as a result of minimizing its potential. Let us see how this result arises in the context of a scalar field in $2+1$ dimensions. This can later be generalized to any dimensions. The action for this field can be written as

\be
S=\frac{1}{g^2}\int d^2 z\left(\pd_z\phi\pd_{\bar z}\phi+V(\phi)\right).
\ee

\noindent We can rescale the coordinates $(z,\bar z)\rightarrow (z\sqrt{\theta},\bar z\sqrt{\theta})$ so the star product depends no longer in $\theta$ and all the dependence in $\theta$ will be in the potential term, which now reads as $\theta V(\phi)$. This implies that in the large noncommutativity region the only relevant term will be the potential.

We will consider a cubic potential

\be
V(\phi)=V_0+m^2\phi\star\phi+\lambda\phi\star\phi\star\phi,
\ee

\noindent then the extremal condition for the energy will be

\be
\label{ppta}
m^2\phi=\lambda\phi\star\phi.
\ee

The explicit form of the solution of this equation is not important for us at this point although yes its meaning. There is an application that allows us to map a $C^\infty$ function into an operator acting on some Hilbert space (could be the single particle Hilbert space). Then we may think of these functions as operators and for our purposes we will use the Weyl prescription

\be
O_{f}(\hat p,\hat q)=\frac{1}{(2\pi)^2}\int d^2k\tilde f(k)e^{-i(k_q\hat q-k_p,\hat p)},
\ee

\noindent where $\tilde f$ is the Fourier transformed of $f(p,q)$. Now the star produts can be seen as usual operator multiplication 

\be
O_f\cdot O_g=O_{f\star g}.
\ee

Now, in the language of operators, equation (\ref{ppta}) is the equation for a projection operator in this Hilbert space. Then the most general solution reads

\be
O=\sum_ia_iP_i,
\ee

\noindent where $a_i$ take values in the set of real extrema of $V(x)$ and $P_i$ are mutually orthogonal projection operators onto one dimensional subspaces \cite{Gop.Min.Str}. 

All this construction sets the tachyon field in the presence of the B-field as a projection operator. Moreover, since it is a real field it is a self-adjoint operator, which leads us back to (\ref{tkgfo}), allowing us to interpret the tachyon as a Fredholm operator. These statements agree with the arguments of sections \ref{ss41} and \ref{ss42} and we can now pursue a interpretation in terms of twisted K-theory, \cite{Har.Moo, Sza, Wit.4}, by considering that the tachyon projects onto $N$-dimensional subspaces so that it represents the map

\be
T\,:\, X\longrightarrow BU(N),
\ee

\noindent into the classifying space of $U(N)$ bundles.

A very interesting feature of noncommutative gauge theories is that in this large commutative region, the noncommutative algebra ${\cal{A}}$ factorizes \cite{Wit.st}. This can be seen in the context of string field theory, where ${\cal{A}}$ is the algebra built by multiplication of string fields. Then, let us write ${\cal{A}}={\cal{A}}_0\otimes{\cal{A}}_1$. This spliting says that the algebra contains two commuting factors, one acting on the string center of mass only and an algebra acting in all the other degrees of freedom. In other words, the algebra ${\cal{A}}_1$ is generated by vertex operators acting on the noncommuting directions and ${\cal{A}}_0$ can carry momentum but only in the commutative directions.

This factorization allows us to give a physical interpretation of Bott's periodicity (\ref{pdb}). Let us suppose, \cite{Har.Moo}, that the space-time manifold splits as $X\times R^2$, where $R^2$ is the noncomutative plane, then we can write this algebra as ${\cal{A}}_1=C(X)\otimes C_{nc}(R^2)$, i.e. the product of the commutative algebra of functions on $X$ and the noncommutative algebra of functions in $R^2$. On the other hand, if one takes the limit in which the B-field vanishes, this algebra will be the commutative algebra of functions in $X\times R^2$, ${\cal{A}}_1=C(X\times R^2)$.

In the K-theory context, the previous argument can be written in the following chain of equalities

\be
K(C(X))=K(C(X)\otimes\KK)=K(C(X)\otimes C_0(R^2)),
\ee

\noindent where the first equality is implied by Morita equivalence and an inductive $N\rightarrow\infty$ limit and the second one is the Weyl map resulting on Bott's periodicity.

%%%%%%%%%%%%%%%%%%%%%%%%%%%%%%%%%%%%%%%%%%%%%%%%%%%%%%%%%%%%%%%
%%%%%%%%%%%%%%%%%%%%%%%%%%%%%%%%%%%%%%%%%%%%%%%%%%%%%%%%%%%%%%%
\subsection{Tachyon condensation and K-homology}\label{ss45}

Let us now, following \cite{Wit.st}, apply the set of ideas devoloped above to the process of tachyon condensation. We will firstly concentrate on type IIA and later on type IIB.

\paragraph{{\sl{type IIA \\[0.5cm]}}}

A key point for this section is the splitting of the algebra explained in section \ref{ss43} and the fact that the algebra ${\cal{A}}_0$ carries all the stringy structure needed here. We can consider two different solutions, consisting in two different vacua, $0$ and $A_0$, the open and closed string vacua respectively, i.e. the first one is understood as the vacuum where a stable D-brane lives after tachyon condensation while the second is the condensation of an elementary configuration. 

Therefore, in the open string sector we find $0-0$, $0-A_0$ and $A_0-A_0$ strings, and we will find physical excitations only in the $0-0$ sector, which allows us to set the condensation of an elementary state. In order to include the effects of the noncommutativity, let us look for a moment at the equation of motion in the bosonic cubic string field theory \cite{Wit.st}

\be
\label{emcsft}
QA+A*A=0,
\ee

\noindent where $Q$ is the BRST operator and $A$ is a string field of +1 ghost number, taken as an element of ${\cal{A}}$. The generalization to the supersymmetric case is not trivial since in this case the string field carries a picture, i.e. something that allows us to eliminate the infinite degeneracy of the BRST-invariant vertex. In terms of the $(\b,\g)$ ghost system, we can write the picture-rainsing operator as $Z=\left\{ Q,\xi\right\}$, where $\xi$ is defined as $\b=e^{-\phi}\pd\xi$. The problem with the direct generalization of the cubic string field theory action is that there is a lack of gauge invariance because of contact-term divergences appearing when two raising-picture operators coincide,\footnote{I am grateful for conversations with P. Resco concerning this subject and for pointing me reference \cite{Ber.1}} leading to a nonpolynomical equation of motion. However, for the purposes of this section, we will deal with equation (\ref{emcsft}).

By virtue of the projector nature of the tachyon field, a solution to (\ref{emcsft}) can be written as $A=A_0\otimes(1-T)$ \cite{Wit.5}. Supose now that this tachyon projects onto some subspace $V$ of $\HS$ such that $\left.T\right|_V=1$ while vanishing in its orthogonal complement $W$, $\left.T\right|_W=0$, then we have again different types of open strings, namely $V-V$, $V-W$ and $W-W$ and again only one of them contains physical excitations, $V-V$. 

As we are working with a torsion B-field, the structure is described in terms of Azumaya algebras. Therefore, the algebraic structure can be written as $A_0\otimes M_m$, with $M_m$ the algebra of matrices $m\times m$. The physical interpretation of this algebra is that it is describing the condensation into a system of $m$ $D(9-2p)$-branes, where $2p$ is the number of noncommutative directions.

\paragraph{{\sl{On the $\ZZ_2$ symmetry\\[0.5cm]}}}

Now that we have seen how K-theory works with and without B-fields, we can explore in more detail the problem of the $\ZZ_2$ symmetry that appears in the vacuum manifold. As we mentioned in section \ref{ss22}, this symmetry appears as a consequence of the spectrum of the tachyon field and denotes two different states that differ in one unit of RR 0-form $G_0/2\pi$ \cite{Hor}. In the D6-brane case this is seen in terms of the two non-trivial complex line bundles defined at infinity by each $U(1)$ factor, which in turn implies that the tachyon field is not homotopic to a constant at infinity \cite{Wit.4}.

A possible solution to this problem was suggested in \cite{Wit.4} by proposing that the gauge group in the world-volume of the initial D9-branes should be $U(\HS)$, which is homotopically trivial in the strong operator topology by virtue of a Kuiper's theorem \cite{Kui}. However, the $N\rightarrow\infty$ limit of \cite{Wit.4} would produce $U(\infty)$, which is not equal to $U(\HS)$, indeed, by a theorem of Palais $U(\infty)$ has the homotopy of $U_{cpt}$, so that all of its odd homotopy groups are nonvanishing and equal to $\ZZ$, \cite{Pal, Har}. Thus it is not clear that an infinite number of D9-branes gives an answer to the question.

However, we can modify the argument by considering non-torsion B-fields. In this case, we argued that, locally, the vacuum manifold could be aproximated by (\ref{vmwntbf}). Therefore, we also have an infinite number of D9-branes, but the quotient by $PU(\HS)$ can be interpreted by saying that the gauge transformations of $PU(\HS)$ are the responsible for the solitonic solutions of type IIA. To extend this construction globally one may found topological obstructions, as those in \cite{Ati.Pat.Sin, Fre.Wit}. Although we do not have a rigurous proof of it, one can argue, see \cite{Gom.Man.3}, that as $PU(\HS)$ is homotopically equivalent to $LE_8$, up to dimension 14, and the global obstruction can be associated with the charge of the D0-branes, we should consider the centrally extended $\widehat{LE}_8$ instead (see \cite{e8} for other arguments on this group) and the vacuum manifold would be (topologically)

\be
{\cal{V}}_{IIA}=\frac{U_{cpt}}{\widehat{LE}_8}
\ee

\paragraph{{\sl{type IIB \\[0.5cm]}}}

The definition of $K^0(X,[H])$ made in section \ref{ss42} is done in terms of projection operators, $\Lambda_\a$, such that $\Lambda^*\Lambda$ and $\Lambda\Lambda^*$ do not commute. Moreover, they define different gauge bundles and the equivalence of them is precisely the defining property of $K^0(X,[H])$. These kind of operators was used in \cite{Har.Moo} in order to a noncommutative ABS construction, as they can be seeing to define a Toeplitz algebra. 

For type IIB we do not have this $\ZZ_2$ symmetry, since we end up with a $U(N)$ gauge group. However there are still some uncertainities remain unsolved. We mentioned in section (\ref{ss21}) that, in type IIB, due to the GSO projection, we can write the string state as

\be
A=\pmatrix{ \nabla & T \cr \bar T & \nabla},
\ee

\noindent now considered as a string field. Let us now write the algebra as

\be
A=\pmatrix{ \a & \b \cr \bar\b & \g}.
\ee

\noindent We can again set $A=A_0\otimes T$ so that, in order to staisfy the equation of motion (\ref{emcsft}), the tachyon must satisfy \cite{Wit.5, Har.Moo}

\be
T\bar T T=T,\qquad \bar T T\bar T=\bar T.
\ee

The case with $T$ invertible is trivial. However, when it is not invertible, we are led to an index theorem as follows. Take $V$ and $W$ of dimensions $n$ and $m$ as the kernels of $T$ and $\bar T$ respectively and $2p$ noncommutative directions. Then the $V-V$ states will be $n$ $D(9-2p)$-branes, $W-W$ states will be $m$ $\overline{D(9-2p)}$-branes and $V-W$ states will be the usual $D(9-2p)$-$\overline{D(9-2p)}$, that decay to the vaccum closed string vacuum. Therefore, this configurations describes a state with charge equal to $n-m$, which is the index of $T$. We can write down an explicit expression for $T$ and $\bar T$, for example

\be
\bar T=\frac{\Gamma_i x^i}{\sqrt{\Gamma_j x^j\bar\Gamma_k x^k}},
\ee

\noindent which resembles equation (\ref{tfftdb}).

\paragraph{{\sl{K-homology \\[0.5cm]}}}

All the previous arguments imply a very important question and it is that the tachyon field is a Fredholm operator (self-adjoint for type IIA). Therefore, we can propose another definition of D-branes and it is that D-branes are Fredholm modules \cite{Asa.Sug.Ter, Sza, Gom.Man.3}.

In a formal language, we see D-branes as the representation spaces of the C$^*$-algebra of Fredholm operators. This proposal has very deep consequences, since these Fredholm modules are the building blocks K-homology, the dual theory of K-theory.

Indeed, this is a natural proposal since K-homology classifies both, cycles and gauge bundles. Therefore, from its structure we can describe both, the gauge bundles on D-branes and the cycles where they can wrap in. Moreover, this presents the advantage of giving physical support for K-homology in the cancellation of the Freed-Witten anomaly.

The role played by K-homology was first proposed in \cite{Per.1} (see also \cite{Van, Ber.Jej.Lei, Lec.Pop.Sza} for different contexts in which enters K-homology) in order to explain the calculation of the phase factors in \cite{Dia.Moo.Wit} in the derivation of K-theory form M-theory, putting further evidence in the relation between $E_8$ bundles and the constructions presented in this paper. Anyway, this subject is out of the scope of this review and we refer the reader to the bibliography.

%%%%%%%%%%%%%%%%%%%%%%%%%%%%%%%%%%%%%%%%%%%%%%%%%%%%%%%%%%%%%%%%%%%%%%%%%%%%%%%%%%%%%%%%%%%%%%%%%%%%%%%%%%%%%%%%%%%%%%%%%%%%%%%%
%%%%%%%%%%%%%%%%%%%%%%%%%%%%%%%%%%%%%%%%%%%%%%%%%%%%%%%%%%%%%%%%%%%%%%%%%%%%%%%%%%%%%%%%%%%%%%%%%%%%%%%%%%%%%%%%%%%%%%%%%%%%%%%%
\section{The D6-brane}\label{s6}

The previous sections allowed us to define the mathematical tools and the structure underlying K-theory and B-fields, concentrating on the non-commutative deformations that can appear in the presence of the B-field. One of the predictions made was that the D6-brane is a 't Hooft-Polyakov monopole for a $U(2)$ gauge theory broken down to $U(1)\times U(1)$. This section is devoted to the study of this soliton mostly from the point of view of SUGRA, in order to see how to reconcile both descriptions. 

Non-abelian monopoles can have an electric charge and these will be the degrees we will identify. This makes a difference with the usual statement concerning the D6-brane, where it could be seen as a Dirac monopole. Our first task will be the definition of the general properties of dyons in field theory. Then we will see how can the dyonic degrees of freedom be interpreted in the context of K-theory and string theory. Finally we will focus on the KK-monopoles. The ten dimensional KK-monopole is known to be dyonic as a prediction of S-duality and carries as electric charge the charge of the H-monopole, i.e. the charge of the B-field. The eleven dimensional KK-monopole, seen as a D6-brane will be shown to be dyonic but now carrying an electric charge associated with the eleven dimensional 3-form, $C^{(3)}$, we will say that it carries one unit of C-monopole charge. 

The fact that the electric charge of the D6-brane can be related to the eleven dimensional 3-form can be traced back to the question of the existence of $E_8$ bundles in eleven dimensions \cite{Dia.Moo.Wit, Gom.Man.2, e8}. However, we will not explore in detail the conditions under which this gauge group appears, instead we will concentrate on finding out which is the gauge group dictated by the SUGRA construction. We will see that both description do not match completely, since from SUGRA we can only obtain an $SU(2)$, instead of the whole $U(2)$ of K-theory.

%%%%%%%%%%%%%%%%%%%%%%%%%%%%%%%%%%%%%%%%%%%%%%%%%%%%%%%%%%%%%%%%%%%%%
%%%%%%%%%%%%%%%%%%%%%%%%%%%%%%%%%%%%%%%%%%%%%%%%%%%%%%%%%%%%%%%%%%%%%
\subsection{Dyons in Field Theory}\label{ss61}

The simplest possible definition of a dyon is that of a state with both, electric and magnetic charges \cite{dyon}. In order to gain a better understanding on the nature of dyons, let us firstly describe what a monopole in field theories is and how the electric charge appears.

From the construction of the Dirac monopole \cite{dira}, we learn an important lesson and it is that the existence of magnetic monopoles implies a requirement of compacity on the $U(1)$ gauge group. As this is not the case in ordinary QED, we must look for a gauge group which contains such a subgroup, for example, $SU(2)$. The monopole constructed in this way is the so-called 't Hooft-Polyakov monopole \cite{thpol}.

Let us consider the Yang-Mills-Higgs lagrangian

\be
{\cal{L}}=-\frac{1}{4}F^a_{\m\n}F^{a\m\n}+\frac{1}{2}D^\m\Phi^aD_\m\Phi^a-V(\Phi),
\ee

\noindent where $F^a_{\m\n}=\pd_\m A_\n^a-\pd_\n A_\m^a+g\e^{abc}A_\m^bA_\n^c$ and $D_\m\Phi^a=\pd_\m\Phi^a+g\e^{abc}A_\m^b\Phi^c$. In the Georgi-Glashow model the potential is

\be
\label{spggm}
V(\Phi)=\frac{1}{4}\lambda\left(\Phi^a\Phi^a-v^2\right)^2.
\ee

Now we can define the Higgs vacuum manifold as

\be
{\cal{V}}=\left\{\Phi^a\;|\; V(\Phi)=0\right\},
\ee

\noindent which, from (\ref{spggm}), is clearly seen to correspond to a $S^2$ with radius $v$.

The finite action condition implies that at infinity the fields take values in this vacuum manifold, so the Higgs field defines the following map:

\be
\label{hom.mon}
\Phi:\;S^2_\infty\longrightarrow {\cal{V}}.
\ee

\noindent These kind of maps are classified by the second homotopy group $\p_2({\cal{V}})$, and are characterized by its degree, i.e. by its winding number:

\be
{\cal{W}}=\frac{1}{4\pi v^3}\int_{S^2_\infty}\frac{1}{2}\e_{ijk}\e^{abc}\Phi^a\pd^j\Phi^b\pd^k\Phi^cdS^i.
\ee

Additionally, the conditions for finite energy, which can be addressed in terms of the condition $T^{00}\geq 0$ in the $00$ component of the energy-momentum tensor imply that $F^{a\m\n}=D^\m\Phi^a=V(\Phi)=0$. A classical vacuum configuration which solves these constraints is given by

\bea
\Phi^a=v\d^{a3},\qquad A_\m^a=0.
\eea

\noindent This constant Higgs field breaks the $SU(2)$ gauge symmetry at infinity down to a $U(1)$. This breaking implies that at the core of the monopole there are spin one charged fields $W^\pm$ with mass $gv$, a massless photon and a Higgs field of mass $\sqrt{2\lambda}v$.

Now we can solve the condition $D_\m\Phi^a=\pd_\m\Phi^a+g\e^{abc}A_\m^b\Phi^c=0$ to get the gauge field

\be
A_\m^a=-\frac{1}{gv^2}\e^{abc}\Phi^b\pd_\m\Phi^c+\frac{1}{v}\Phi^af_\m,
\ee

\noindent which leads to a non-abelian field strength

\be
F^{a\m\n}=\frac{1}{v}\Phi^aF^{\m\n}=\frac{1}{v}\Phi^a\left(-\frac{1}{gv^3}\e^{bcd}\Phi^b\pd^\m\Phi^c\pd^\n\Phi^d+\pd^\m f^\n-\pd^\n f^\m\right).
\ee

The equations of motion for this system are

\bea
D_\m F^{a\m\n} & = & g\e^{abc}\Phi^b D^\n\Phi^c, \\
\left( D_\m D^\m\Phi\right)^a & = & -\lambda\Phi^a\left(\Phi^b\Phi^b-v^2\right),
\eea

\noindent plus a Bianchi identity $D_\m{}^*F^{a\m\n}=0$. These conditions imply that

\bea
\pd_\m F^{\m\n} & = & 0, \\
\pd_\m{}^*F^{\m\n} & = & 0.
\eea

This set of equations imply that outside the core of the monopole the non-abelian gauge field is a pure gauge in the direction of $\Phi^a$.

The magnetic charge, $m$, of this monopole can be computed using Gauss' law for the magnetic field integrated over the sphere at infinity, which implies a Dirac quantization condition in terms of the winding number

\be
gm=4\p{\cal{W}}.
\ee

On the other hand, we can see that this magnetic charge is related to the mass of the soliton. This relation is the Bogomol'nyi bound:

\bea
\nonumber M & = & \int d^3r\left(\frac{1}{2}\left(\vec B^a\cdot\vec B^a+\vec D\Phi^a\cdot\vec D\Phi^a\right)+V(\Phi)\right)\geq\int d^3r\frac{1}{2}\left(\vec B^a\cdot\vec B^a+\vec D\Phi^a\cdot\vec D\Phi^a\right)\\
& = & \frac{1}{2}\int d^3r\left(\vec B^a-\vec D\Phi^a\right)\cdot \left(\vec B^a-\vec D\Phi^a\right)+vm.
\eea

\noindent Therefore, the bound is

\be
M\geq vm,
\ee

\noindent where the equality occurs when the potential vanishes, $V(\Phi)=0$, and the Higgs and magnetic fields satisfy the Bogomol'nyi equation

\be
\label{bgeq}
\vec B^a=\vec D\Phi^a,
\ee

\noindent which implies that we can consider the Higgs field as the scalar potential for the magnetic field \cite{bps}.

In order to find an explicit solution, one can make the following simple ansatz:

\be
\label{ansazt.1}
\matrix{\Phi^a & = & \frac{\hat r^a}{gr}H(vgr), \cr
&&\cr
A_i^a & = & -\e^a{}_{ij}\frac{\hat r^j}{gr}\left( 1-K(vgr)\right),}
\ee 

\noindent which has the required symmetry by the boundary conditions at infinity, where the fields are invariant under the diagonal $SO(3)$ of $SO(3)_G\times SO(3)_R$, generated by $\vec K=\vec J+\vec T$, of global gauge transformations plus rotations. When it is substituted in (\ref{bgeq}), produces the following solutions:

\bea
\label{sol.1}
H(x) & = & x\coth x-1,\\
\label{sol.2}
K(x) & = & \frac{x}{\sinh x},
\eea

\noindent where $x=vgr$. 

The topological properties we have described and the set of equations (\ref{ansazt.1})-(\ref{sol.2}) can be compared with the K-theoretical solution for the D6-brane (\ref{sktds}). This is what allow us to say that the D6-brane is a non-abelian monopole. However, there is a subtlety concerning the gauge groups. While in this case we have an $SU(2)$ gauge theory, the D6-brane needs a $U(2)$, therefore in K-theory there is an extra $U(1)$. This question is focused in \cite{Gom.Man.3}.

The study of the electric charge of the monopole is the main purpose of this section, thus let us stop a second in order to clarify some ideas. On one hand we have the Julia-Zee dyon, \cite{dyon}, where the electric charge is coupled to the electric field, which is a pure gauge field non-vanishing at infinity. On the other hand we find Witten's dyon effect, \cite{Wit.6}, where the electric charge couples to the magnetic field due to the presence of the $\theta$ term in the lagrangian.

The Julia-Zee electric charge appears when we introduce in the ansazt (\ref{ansazt.1}) the temporal component of the gauge field as

\be
A_0^a=\frac{\hat r^a}{gr}J(vgr).
\ee

\noindent Then the electric part of the field strength, $F^a_{0i}$, is non-vanishing and we can apply Gauss' law to find the electric charge.

In the Witten's effect, we add a topological coupling to the Yang-Mills-Higgs lagrangian which breaks $CP$-invariance

\be
{\cal{L}}=-\frac{1}{4}F^a_{\m\n}F^{a\m\n}+\frac{\theta g^2}{32\p^2}F^a_{\m\n}{}^*F^{a\m\n}+\frac{1}{2}D^\m\Phi^aD_\m\Phi^a-V(\Phi),
\ee

\noindent we can now apply a a gauge transformation in the direction of the unbroken $U(1)$, i.e. in the direction of $\Phi^a$:

\bea
\d A_\m^a & = & \frac{1}{vg}D_\m\Phi^a,\\
\d\Phi^a & = & 0,
\eea

\noindent then we can apply the Noether's method and find that the conserved charge associated to this transformation is

\be
N=\frac{e}{g}+\frac{\theta gm}{8\p^2},
\ee

\noindent where $(e,m)$ are the electric and magnetic charges respectively. Requiring $U(1)$ invariance $e^{2\p iN}=1$ we find that the electric charge is

\be
\label{cedmdtp}
e=ng-\frac{\theta g^2}{8\p^2}m=ng-\frac{\theta{\cal{W}}g}{2\p},
\ee

\noindent where we have taken into account the relation between the magnetic charge and the winding number. Equation (\ref{cedmdtp}) implies the important result that for $\theta\neq 0$ there are no electrically neutral monopoles.

%%%%%%%%%%%%%%%%%%%%%%%%%%%%%%%%%%%%%%%%%%%%%%%%%%%%%%%%%%%%%%%%%%%%%
%%%%%%%%%%%%%%%%%%%%%%%%%%%%%%%%%%%%%%%%%%%%%%%%%%%%%%%%%%%%%%%%%%%%%
\subsection{Dyons and K-theory}\label{ss62}

In the moduli space fo monopoles, the electric charge appears as another coordinate. In particular, the charge one monopole moduli space of monopoles with fixed centre is

\be
{\cal{M}}=\RR^3\times S^1,
\ee

\noindent where the $\RR^3$ piece corresponds to the group of translations and the $S^1$ corresponds to the fibration of the global $U(1)$ gauge group, so that motions along this fibre generate the elecrtic charge. This extra $S^1$ fibration also produces the desingularization of the moduli space.

A usual mechanism in string theory is the promotion of gauge symetries to space-time simetries of some internal manifold. Therefore, the proposal of this section is promoting the $S^1$ fiber to a space-time coordinate \cite{Gom, Gom.Man.2}. Then this moduli space resembles the construction of section \ref{sss251} for $K^{-1}$.

Let us take the K-group as $K(\RR^3\times S^1)=K_{cpt}(S^3\times S^1)$. Then, from section \ref{ss22}, this is the same as $K^{-1}(S^3)$. This is precisely the relevant K-group for the D6-brane and, moreover, it is associated with a $\pi_3$. This homotopy group is important since, in the presence of a non-vanishing $\theta$, the gauge transformation defining the electric charge has a non-trivial winding number in it.

The message of this construction is that the existence of an electric charge for the D6-brane should be considered as a trace of the eleventh dimension.

%%%%%%%%%%%%%%%%%%%%%%%%%%%%%%%%%%%%%%%%%%%%%%%%%%%%%%%%%%%%%%%%%%%%%
%%%%%%%%%%%%%%%%%%%%%%%%%%%%%%%%%%%%%%%%%%%%%%%%%%%%%%%%%%%%%%%%%%%%%
\subsection{Dyons in String Theory}\label{ss63}

In string theory we can also find dyonic states, now carrying (electric) winding and (magnetic) KK charges. In this case, dyons exist as a prediction of S-duality and must have the same degeneracy as a purely winding state \cite{Sen.3}.

Let us take the vector

\be
\vec\a=\pmatrix{p\cr\omega},
\ee

\noindent where $p$ denotes the momentum and $\omega$ the winding along the compact direction of spacetime. In turn, we can take

\be
L\vec\b=\pmatrix{ 0 & 1 \cr 1 & 0}\vec\b=\pmatrix{n_k\cr n_H},
\ee

\noindent where $n_k$ is the KK-charge, $n_H$ is the H-monopole charge and $L$ is the metric on the momentum and winding lattice along the compact direction. 

S-duality acts on these states by means of an $SL(2,\ZZ)$ matrix as

\be
\pmatrix{\vec\a\cr\vec\b}\longrightarrow\pmatrix{a & b \cr c & d}\pmatrix{\vec\a\cr\vec\b},
\ee

\noindent where $ad-bc=1$ and $a,b,c,d\in\ZZ$. 

We can take an initial state with one unit of winding charge, the S-duality produces:

\be
\left\{\vec\a=\pmatrix{0\cr 1},\vec\b=\pmatrix{0\cr 0}\right\}\stackrel{S}{\longrightarrow}\left\{\vec\a=\pmatrix{0\cr a},\vec\b=\pmatrix{0\cr c}\right\}.
\ee

Therefore, S-duality implies that it is dual to a state carrying electric-winding and magnetic-Kaluza-Klein charges.

The analogy with the field theoretic dyons can be seen as follows. Take the state $(n,{\cal{W}})=(1,0)$, which corresponds to a $W^+$, then we have

\be
\pmatrix{1\cr 0}\longrightarrow\pmatrix{a & b \cr c & d}\pmatrix{1\cr 0}=\pmatrix{a\cr c},
\ee

\noindent so this duality requires the existence of states with $(n,{\cal{W}})=(a,c)$, which for $c=1$ are the dyonic excitations corresponding to the charge 1 BPS monopole of the previous section.

%%%%%%%%%%%%%%%%%%%%%%%%%%%%%%%%%%%%%%%%%%%%%%%%%%%%%%%%%%%%%%%%%%%%%
%%%%%%%%%%%%%%%%%%%%%%%%%%%%%%%%%%%%%%%%%%%%%%%%%%%%%%%%%%%%%%%%%%%%%
\subsubsection{KK-monopoles}\label{sss631}

A solution in string theory that possesses naturally a KK-charge is the KK-monopole, that takes the form \cite{Sor, Gro.Per}

\be
ds^2=dy^2+ds^2_{TN},
\ee

\noindent with $ds^2_{TN}$ the self-dual Taub-Nut metric \cite{Gib.Haw}

\be
\label{taubnut}
ds^2_{TN}=U\left[ dx_4+4m\left(1-\cos\theta\right)d\phi\right]^2+U^{-1}\left[dr^2+r^2\left(d\theta^2+\sin^2\theta d\phi^2\right)\right],
\ee

\noindent where $x^4$ coordinate is known as the Taub-NUT coordinate and is periodic of period $2\p$ due to the fact that it represents the isometric direction, but this has also one advantage, and is that this periodicity cancels the Dirac singularity of the metric, reflecting the fact that it is neccesary an isometric direction for the existence of this solution, and $U$ is an harmonic function in Taub-NUT satisfying the following boundary conditions

\be
U^{-1}\rightarrow\left\{
\matrix{ \sim \frac{1}{r},& |x| & \rightarrow & 0, \cr 1, & |x| & \rightarrow & \infty,}
\right.
\ee

\noindent and can be taken to be

\be
U=\left( 1+\frac{4m}{r}\right)^{-1}
\ee

The set of bosonic zero modes of the KK-monopole is not big enough to contain the dyonic degree of freedom. However, Taub-NUT we have an extra zero mode associated with pure gauge transformations of the B-field that are non-vanishing at infinity. This 2-form can be taken as, \cite{Sen.3}

\be
\label{pgf}
B=\a d\left(\frac{r}{r+4m}\s_3\right),
\ee

\noindent with $\s_3=\frac{1}{4m}\left[dx_4+4m(1-\cos\theta)d\phi\right]$. Obviously, the pure gauge field (\ref{pgf}) is non vanishing at $r\rightarrow\infty$. Moreover, it is a zero mode since it is proportional to the unique harmonic two form in euclidean Taub-Nut space.

In references \cite{Sen.3, Greg.Har.Moo}, it was pointed out that the previous derivation of the dyonic charge of KK-monopoles in string theory is very much similar to the characterization of the moduli space of BPS-monopoles in $SU(2)$ Yang-Mills-Higgs gauge theories. In this case, the pure $U(1)$ gauge transformation $e^{i\a\Phi(\vec x)/|\Phi(\vec x)|}$, with $\Phi$ the Higgs field and $\a\in[0,2\pi]$ is non vanishing at infinity in the monopole background. The parameter $\a$ defines the coordinate on the $S^1$ fiber of the moduli space ${\cal{M}}_1={\mathbb{R}}^3\times S^1$ and the motion along this fiber generates the electric charge of the monopole that is defined by the corresponding conjugate momentum.

In the KK case we observe that continous changes of $\a$ in (\ref{pgf}), i.e. a one parameter family of type $B(t)=\a_0td\left(\frac{r}{r+4m}\s_3\right)$ generates a non vanishing $H$ field with winding charge proportional to the conjugate momentum $\a_0$ \cite{Sen.3}.

The analogy between the dyon effect for the KK-monopoles in string theory and the dyonic charge for BPS-monopoles can be pushed a bit further. For instance, in the BPS case for the 't Hooft-Polyakov $SU(2)$ monopole, the relevant homotopy groups are

\be
\pi_3\left(SU(2)\right)=\pi_2\left(SU(2)/U(1)\right)=\pi_1\left(U(1)\right)={\mathbb{Z}}.
\ee

The gauge transformation $e^{i\a\Phi(\vec x)/|\Phi(\vec x)|}$ we have used to generate the dyon zero mode is for $\a=2\pi$ a non trivial gauge transformation in $\pi_3\left(SU(2)\right)$ with winding number equal to minus the magnetic charge of the monopole. as we have seen in section \ref{ss61}, this fact is crucial in order to derive Witten's relation between the dyon electric charge and the instanton $\Theta$ vacuum angle \cite{Wit.6}.

In this case, the role of this $\pi_3$ is played by the winding

\be
\label{wn}
\frac{1}{16\pi^2}\int_{S^3}\omega,
\ee

\noindent for $S^3$ the boundary defined by the Hopf fibration and $\omega$ the totally anti-symmetric part of the spin connection \cite{Hul}. This winding is known as the K-charge of Taub-NUT and is proportional to the NUT charge of the gravitational instanton. This spin connection can be written as $F\wedge k$ where $F$ is the field strength of the KK vector field and $k$ is the Killing vector along the compact direction. 

It is interesting to relate this result with the heterotic KK-monopole. In this case the winding is given in terms of the integral of the Lorentz-Chern-Simons form which contributes, by anomaly cancellation arguments \cite{Gre.Sch}, to the field strength of the antisymmetric tensor field $B_{\m\n}$ and (\ref{wn}) becomes proportional to the H-charge $\int_{S^3}H$. Moreover, the H-field defined by $\omega$ is equal to $K\wedge B$ for $B$ given by (\ref{pgf}) with $\a=2\pi$ and where $\int_{S^1}K=2\pi R$ is the lenght of the $S^1$ fiber at infinity.

Therefore, notice that the winding number (\ref{wn}) only depends on the Taub-Nut geometry and is the same for all string KK-monopoles, and, as we will see in sections \ref{sss652} and \ref{ss66} how it can be related to the antisymmetric tensor B-field and its gauge transformations used to define the dyon zero mode. The relation with the cancellation of anomalies is explored in \cite{Gom.Man.3}

The physics underlying the process can be stated as follows \cite{Greg.Har.Moo, Gib.Rub}. In ten dimensions a KK-monopole metric is given by ${\mathbb{R}}^{5,1}\times{\cal{N}}_4$, where ${\cal{N}}_4$ is the self-dual euclidean  Taub-NUT space. The boundary of ${\cal{N}}_4$ at spatial infinity is the 3-sphere, $S^3$, interpreted as the Hopf fibration of $S^1$ on $S^2$. In string theory we can consider winding states on $S^1$ which, due to the fact that $\pi_1(S^3)=0$, can unwind in the presence of a KK-monopole. Conservation of winding charge implies that the KK-monopole must carry winding charge\footnote{For a K-theory discussion of this phenomena, see reference \cite{Mal.Moo.Sei.1}}. After standard KK reduction this winding charge becomes electric charge for the $B_{\m 4}$ $(\m=1,2,3)$ component of the B-field along the compact $S^1$ direction.

%%%%%%%%%%%%%%%%%%%%%%%%%%%%%%%%%%%%%%%%%%%%%%%%%%%%%%%%%%%%%
%%%%%%%%%%%%%%%%%%%%%%%%%%%%%%%%%%%%%%%%%%%%%%%%%%%%%%%%%%%%%

\subsection{D6-branes and KK-monopoles}\label{ss64}

Now that we have explored the stringy KK-monopole, we can proceed to our next step, which is the study of the eleven dimensional KK-monopole and its ten dimensional counterpart, the D6-brane. In order to characterize properly the solution, let us start with the 11-dimensional superalgebra \cite{Hol.Pro}

\bea
\label{edsa}
\nonumber \left\{Q_\a,Q_\b\right\} & = & \left(\Gamma^MC^{-1}\right)_{\a\b}P_M+\frac{1}{2!}\left(\Gamma^{MN}C^{-1}\right)_{\a\b}Z_{MN}+\\
& & +\frac{1}{5!}\left(\Gamma^{MNPQR}C^{-1}\right)_{\a\b}Z_{MNPQR},
\eea

\noindent where due to the fact that the non-trivial representation of the Clifford algebra is inherited from the one in ten dimensions, its dimension is $2^{(D-1)/2}=2^{10/2}=32$, so $Q_\a$ is a Majorana spinor with 32 real components, $C$ is the charge conjugation matrix, $P_M$ is the generator of translations and $Z_{MN}$ and $Z_{MNPQR}$ are the central charges. The indices run over $\a=1,..,32$ and $M=0,..,10$.

The purely spatial components of the central charges of the superalgebra can be associated with extended objects. However, more interesting for us are the temporal components of these central charges \cite{Hul}. In general we can take

\be
\hat Z_{j_1...j_p}=\frac{1}{(q+1)!}\e_{j_1...j_p0i_1...i_q}Z^{0i_1...i_q},
\ee

\noindent which, for the central charges of (\ref{edsa}), can be written as

\bea
\hat Z_{j_1...j_6} & = & \frac{1}{5!}\epsilon_{j_1...j_60i_1...i_4}Z^{0i_1...i_4}, \\
\hat Z_{j_1...j_9} & = & \frac{1}{2!}\epsilon_{j_1...j_90i}Z^{0i},
\eea

\noindent which indicates the existence of some kind of $M6$ and $M9$ branes. In this chapter we are concerned with this M6-branes.

Imposing the condition $\langle\left\{Q_\a,Q_\b\right\}\rangle=0$ we can see how many supersymmetries are preserved. Let us choose a configuration in which only $P_0$ and $Z_{0789\#}$ (where $\#$ indicates the eleven dimension) are different from zero, while all the other bosonic charges vanish. Then

\be
\label{vps}
\left\{ Q_\a,Q_\b\right\}=P_0\left( 1-\Gamma^{789\#}\tilde q_5\right),
\ee

\noindent where $\tilde q_5=Z_{0789\#}/P_0$.

The matrix $\left(\Gamma^{789\#}\right)^2=1$, and so it has two eigenvalues $\pm 1$ and null trace. On the other hand, been the l.h.s of (\ref{vps}) a quadratic form of Majorana spinors, it is always $\geq 0$, so we must have $|\tilde q_5|\leq 1$, which becomes equal to one when we saturate the bound, and then the condition for unbroken supersymmetry becomes

\be
\label{hosds}
\Gamma^{123456}\e=\pm\e.
\ee

\noindent which implies that this M6-brane preserves half of the supersymmetries and should be associated to a 7-form in 11-dimensions. However, here we find a problem, namely, that there is no such a 7-form in $D=11$ SUGRA which could be used as gauge potential for this object, at least on Minkowski space.

Fortunately, this is not the end of the story, because there is a situation in which this form may not be trivial and it is to consider a direction compactified on $S^1$ \cite{Hul, Tow}. When we compactify the eleventh direction, we find

\bea
g_{MN}\stackrel{S^1}{\longrightarrow}\left\{
\matrix{ g_{\m\n} & & \cr g_{\m\#} & \equiv & A_\m \cr g_{\#\#} & \equiv & \phi}
\right.,
\qquad
P_M\stackrel{S^1}{\longrightarrow}\left\{
\matrix{ P_\m & & \cr P_\# & \equiv & Q.}
\right.
\eea

Now we can look for the magnetic dual of $A_\m$ in D-dimensions, which is easily seen to correspond to a (D-5)-form. Performing a dimensional reduction in this form we obtain

\be
K_{i_1...i_{D-5}}\stackrel{S^1}{\longrightarrow}\left\{
\matrix{
P_{j_1...j_{D-5}}, \cr \tilde K_{j_1...j_{D-6},}
}
\right.
\ee

\noindent where the $P_{j_1...j_{D-5}}$ is related to the magnetic charge of the graviphoton and $\tilde K_{j_1...j_{D-6}}$ is what has come to call the K-charge, which is related to the NUT charge.

Summrizing, two gravitational charges in D-dimensions, $(P,K_{D-5})$, produce two different sets of charges when a dimension is compact, namely a pair of gravitational charges $(P,\tilde K_{D-6})$ and a pair of electromagnetic charges $(Q,K_{D-5})$. With the extra implication that this eleven dimensional 6-form has to be non-trivial in the presence of a KK-monopole. Therefore, the metric takes the form

\be
\label{kkmed}
{\cal{N}}_4\times\RR^{D-5,1},
\ee

\noindent where ${\cal{N}}_4$ is the Taub-NUT 4-dimensional self-dual Euclidean space (\ref{taubnut}), where now, the Taub-NUT coordinate is interpreted as the eleventh compact direction.

Now we can perform a direct dimensional reduction of this KK-monopole. It would be interesting to remember that in this kind of dimensional reduction the compact direction is in the transverse space to the world-volume of the branes, which gives rise to the unwrapped branes. Moreover, the tensions remain unchanged, while the harmonic function changes as

\be
{\mathbb{E}}^{10-p}\rightarrow{\mathbb{E}}^{9-p}:\;\int_{-\infty}^\infty\frac{dx^s}{\left[\left( x^s\right)^2+r^2\right]^{(8-p)/2}}\simeq\frac{1}{r^{7-p}}.
\ee

Taking into account the known relation between M-theory parameters $(l_p,R_s)$ and type IIA parameters $(l_s,g_s)$ given by the expressions $\frac{R_s}{l_p}=g_s^{2/3}$ and $l_p^3=g_sl_s^3$, the relation between the coupling constant and the dilaton, $g_s=e^{2\phi}$ and the KK-ansazt for the dimensional reduction

\be
ds_{11}^2=R_s^2\left( dx^s+{\cal{A}}_\m dx^\m\right)^2+ds_{10}^2,
\ee

\noindent we find that the ten dimensional descendant of the eleven dimensional KK-monopole takes the form

\be
\label{dsbm}
ds^2_{10}=H^{-1/2}ds^2\left({\mathbb{E}}^{1,6}\right)+H^{1/2}ds^2\left({\mathbb{E}}^{3}\right),
\ee

\noindent with

\bea
e^{-2\phi}=H^{3/2},\quad H=1+\frac{k}{r},\quad F_m={^*}dH.
\eea

\noindent Then we see that the metric (\ref{dsbm}) corresponds to the supergravity solution of the D6-brane in type IIA.

There is an interesting structure of singularities relating (\ref{kkmed}) and (\ref{dsbm}), namely, while (\ref{dsbm}) is singular when $r=0$, which is interpreted as the location of the brane, (\ref{kkmed}) is a completely non-singular solution of the Einstein equations in eleven dimensions, and the singularity at $r=0$ is simply a coordinate one. To see it, let us take limit $r\rightarrow 0$, $H\sim 1/r$. We can take $\rho=\sqrt{r}$ to get

\be
ds_4^2\sim d\rho^2+\rho^2\left[d\Omega^2+\left(dy+{\cal{A}}\right)^2\right],
\ee

\noindent now, as $x^4\sim x^4+2\p$, the surfaces at $\rho$ constant are 3-spheres, which implies that the metric is asymptotically ${\mathbb{E}}^4$ in spherical coordinates. However, in the limit $r\rightarrow\infty$, the metric takes the form of the KK-vacuum, so this eleven dimensional metric is a gravitational instanton:

\be
\left.
\matrix{ {\mathbb{E}}^{(1,9)}\times S^1 \cr {\mbox{{\sl KK-vacuum}}} \cr r\rightarrow\infty }
\right\}
\longleftrightarrow
\left\{
\matrix{ {\mathbb{E}}^{(1,10)} \cr {\mbox{{\sl M-theory vacuum}}} \cr r\rightarrow 0. }
\right.
\ee

The most important question here is that we are seeing a process of desingularization induced by the eleventh dimensional coordinate, and, what is even more relevant for our purposes, it is produced by the same effect than that for monopoles in field theory, i.e. by the $S^1$ fiber in the moduli space, now promoted to a physical coordinate. 

This process of desingularization is reminiscent of the usual ones described in the framework of the renormalization methods applied to field theory. Namely, we have seen the process of desingularization of an object by going to strong coupling, which in turn has allowed us to uncover some hidden structure. In this case, as we are working with purely gravitational solution, the hidden structure is the own space-time, and a new dimension has appeared.

Another advantage of this process is that now we can put on precise grounds the eleven dimensional origin of the fields in the world-volume of the D6-brane. This can be traced back to the decomposition of the eleven dimensional 3-form of section \ref{ss33}. From there we see that we have a gauge field in (6+1) dimensions and 3 scalars coming from the translational zero modes. This is the bosonic content of a vector multiplet for $N=1$ supersymmetric Yang-Mills in (6+1) dimensions \cite{Hul, Sen.4}, and can be seen as originated by a wrapped membrane with one of its ends in the world volume of the KK-monopole \cite{Ber.Ort.Jan, Gom.Man.1} (see \cite{Spa} for an interpretation in terms of cancellation of global membrane anomalies).

Requiring harmonicity in those zero modes in the decomposition of the 3-form of M-theory that live in the transverse space, we obtain the relation \cite{Ima}

\be\label{G}
\nabla_{[\m}V_{\n]}=\frac{1}{2\pi\a'}C_{\m\n},
\ee

\noindent which can be seen as giving the gauge transformation for $C_{\m\n}$ and establishes the standard relation between the B-field and the gauge field on the D6-brane, as can be seen building the corresponding part of the 4-form field strenght as

\bea
\label{gm}
\nonumber G_{\m\n ij} & =& C_{\m\n}\nabla_{[i}A_{j]}+\nabla_{[\m}A_{\n]}B_{ij}\\
\nonumber & = & C_{\m\n}\left( \nabla_{[i}A_{j]}+\frac{1}{2\pi\a'}B_{ij}\right)\\
& = & C_{\m\n}{\cal F}_{ij}.
\eea 

\noindent When the $C_{\m\n}$ is integrated, gives rise to the $F+B$ term of the Born-Infeld theory of the D6-brane. Moreover, since the holonomy group of Taub-NUT is $SU(2)$, in order to obtain a normalizable $C_{\m\n}$, we must require for it to be an anti-self-dual tensor \cite{Ima}.

There is an extra condition which we have not used, and it is that the field strenght of the 3-form of M-theory, $G$, has to be zero. From (\ref{gm}) one obtains the following relation between the fields on the world-volume

\be
\label{F}
\frac{1}{2\pi\a'}B+dA=0,
\ee 

\noindent which trivially implies that the B-field is flat, i.e. its topological classification lies in $[H]\in Tors\left( H^3(X,{\mathbb{Z}})\right)$ \cite{Gom.Man.1}.

%%%%%%%%%%%%%%%%%%%%%%%%%%%%%%%%%%%%%%%%%%%%%%%%%%%%%%%%%%%%%
%%%%%%%%%%%%%%%%%%%%%%%%%%%%%%%%%%%%%%%%%%%%%%%%%%%%%%%%%%%%%
\subsection{The D6-brane as a dyon}\label{ss65}

Let us now work this analogy between KK-monopoles and BPS-monopoles. The dyonic nature of the D6-brane as a 't Hooft-Polyakov monopole for the open tachyon condensation derives from the fact that the gauge transformation $e^{i\a T\left(\vec x\right)}$ is non vanishing at infinity and for $\a=2\pi$ is topologically non trivial in $\pi_3\left(U(2)\right)$. 

The K-theory description of D6-branes as 't Hooft-Polyakov  $U(2)$ monopoles imply that they can carry dyonic charge. Our goal now will be to understand this dyonic charge from the point of view of D6-branes as M-theory KK-monopoles, when the $S^1$ is Hopf fibered over space-time \cite{Gom.Man.2}.

Similarly to the case of the KK-monopoles in string theory, we should look for some non trivial gauge transformations non vanishing at infinity. In the case of M-theory KK-monopoles, these transformations are going to be gauge transformations related to the 3-form field $C$. However they will not transform the whole 3-form, because there is only one normalizable harmonic form in the Taub-NUT space and it is a 2-form, so we should consider the decomposition of the 3-form in this space. 

In section \ref{sss652} we will see that this dyonic charge is, again, topologically related to a non vanishing winding number in $\pi_3$, which we may write as associated to a 3-form proportional to the volume of the $S^3$ at infinity 

\be
\label{a}
\Omega=\a\omega_3,
\ee

\noindent and, after an appropiated normalization, this winding number is simply $\int_{S^3}\Omega$ for $\a=2\pi$. We can now give a nice gauge theory meaning. In fact, if we assume that the M-theory 3-form $C$ can be written as a Chern-Simons 3-form for a $E_8$ gauge field theory plus the gravitational Chern-Simons term \cite{Dia.Moo.Wit, Hor.Wit, Wit.7,e8} then we can easily relate $\int_{S^3}\Omega$ to the topology of the M-theory $E_8$ vector bundle. In fact, if we work with the strength 4-form $G$ of $C$ and use the relation  \cite{Wit.7}

\be
\left[\frac{G}{2\pi}\right]=w(V)-\frac{\l}{2},
\ee

\noindent where $\l=p_1/2$, $\left[G/2\pi\right]$ denotes the cohomology class of $G/2\pi$ and $w(V)$ is the second Chern class of the $E_8$ vector bundle V, we observe that for a four dimensional euclidean Taub-Nut space, which has vanishing first Pontryagin number, the cohomology class of $G$ is completely determined by the M-theory $E_8$ vector bundle. Thus we are tempted to conjecture that the $\pi_3$ associated with the dyonic properties of the KK-M-theory monopole are directly related with the second Chern class of the formal $E_8$ bundle defined on the M-theory eleven dimensional space-time \cite{Gom.Man.2}.

Let us just point out that it is very natural to associate the heterotic 5-brane with the non vanishing $\pi_3\left( E_8\right)$. However, in this case the gauge theory is part of the closed string spectrum of the heterotic string. Probably a heterotic origin of closed string gauge fields can be a good hint to unravel the deep physical meaning of the gauge theories appearing in the K-theory description of D-branes, in accordance with the exposition made in sections \ref{ss41} and \ref{ss45}.

We have argued that the $S^1$ fiber of the moduli space{\footnote{More generally, the $S^1$ fiber of the multimonopole space ${\mathcal{M}}_k={\mathbb{R}}^3\times\frac{S^1\times {\mathcal{M}}_k^0}{{\mathbb{Z}}_k}$}} of the D6-brane as a 't Hootf-Polyakov monopole would have a nice interpretation in terms of the extra M-theory dimension. However, the D6-brane is not really allowed to move in this direction.

The connection between the M-theory extra dimension and the topology of monopole moduli space can be considered from a different point of view. Namely, if we consider $k$ monopoles, the moduli space

\be
\label{mmms}
{\mathcal{M}}_k={\mathbb{R}}^3\times\frac{S^1\times {\mathcal{M}}_k^0}{{\mathbb{Z}}_k},
\ee

\noindent is such that \cite{Ati.Hit}

\be
\pi_1\left({\mathcal{M}}_k\right)={\mathbb{Z}},
\ee

\noindent which comes from two facts. First, in a particle like approximation we will get $\left({\mathcal{M}}_1\right)^k$, with $\pi_1={\mathbb{Z}}^k$. Secondly, $\pi_1\left({\mathcal{M}}_k^0\right)={\mathbb{Z}}_k$. Thus we get from (\ref{mmms}) $\pi_1\left({\mathcal{M}}_k\right)={\mathbb{Z}}$. For a KK-monopole of charge $k$, the boundary looks like $S^3/{\mathbb{Z}}_k$, with $\pi_1\left( S^3/{\mathbb{Z}}_k\right)={\mathbb{Z}}_k$. On the other hand, for each KK-monopole we have an harmonic 2-form which produces ${\mathbb{Z}}^k$. When we combine the two facts, like in the characterization of $\pi_1\left({\mathcal{M}}_k\right)$, we get ${\mathbb{Z}}$ as a final result (see \cite{Mal.Moo.Sei.1} for a discussion in terms of singletons).

%%%%%%%%%%%%%%%%%%%%%%%%%%%%%%%%%%%%%%%%%%%%%%%%%%%%%%%%%%%%%%%%%%
%%%%%%%%%%%%%%%%%%%%%%%%%%%%%%%%%%%%%%%%%%%%%%%%%%%%%%%%%%%%%%%%%%
\subsubsection{Electric and magnetic charges of the D6-brane}\label{sss652}

Let us now compute the charges of the system, \cite{Gom.Man.1}. Taking (\ref{taubnut}), we can write the KK vector potential and its field strength as

\bea
A^{(1)}&=&4m(1-\cos\theta) d\phi,\\
F&=&dA=4m\sin\theta d\theta\wedge d\phi.
\eea

From here one can compute the NUT charge as 

\be
N=\frac{1}{8\pi}\int F=m,
\ee

\noindent which is set from boundary conditions \cite{Ima} to be $m\propto l_se^{\phi_0}$, where $\phi_0$ is the boundary value of the dilaton of the type IIA string theory.

The magnetic charge of the monopole can be computed from the integral of the totally antisymmetric part of the spin connection minus the background $\omega=F\wedge k$, where $k=\left(1+\frac{4m}{r}\right)^{-1}\left(dx^4+4m(1-\cos\theta) d\phi\right)$ is the Killing vector representing the isometry, which equals

\be
\label{mf}
\omega=4m\sin\theta\frac{r}{r+4m}d\theta\wedge d\phi\wedge dx^4.
\ee

\noindent This form is proportional to the volume form of the $S^3$ which means that it can be interpreted as a winding number. The corresponding charge is \cite{Hul}

\be
K=\frac{1}{16\pi^2}\int \omega=4m^2.
\ee

\noindent which is the quantity playing the role of the $\pi_3$ mentioned in section \ref{sss631}.

The metric (\ref{taubnut}) is self-dual, i.e. $R_{\a\b\g\s}=\frac{1}{2}\e_{\a\b\eta\pi}R^{\eta\pi}{ }_{\g\s}$, which implies that the mass of the solution is equal to its NUT charge, and has a positive orientation \cite{Gib.Haw} defined by an orthonormal frame $e^A=\left\{ e^{x^4},e^a\right\}$ which can be chosen in such a way that

\bea
e^A{}_Me^B{}_N\eta_{AB}=g_{MN},\qquad \hat e^a{}_\m \hat e^b{}_\n\eta_{ab}=\left(\frac{r+4m}{r}\right)g_{\m\n}
\eea

\noindent with 

\bea
\label{votn}
e^{\underline x^4}  & = & \left(\frac{r}{r+4m}\right)^{1/2}\left(dx^4+4m(1-\cos\theta d\phi\right),\\
e^a                 & = & \hat e^a{}_\m dx^\m,
\eea

\noindent where we have $(\underline x^4,a)$ denote the flat coordinates and $(x^4,\m)$ the curved ones.

Let us now compute the effects of the eleven dimensional 3-form. We only need the part of it living in Taub-NUT so, once again, we turn to the decomposition made in section \ref{ss33} and work with the 2-form $C_{\m\n}$ and the 1-form $V_\m$. Using (\ref{G}), we observe that the 2-form $C_{\m\n}$ is a pure gauge. It is this pure gauge 2-form the one that is playing for the eleven dimensional KK-monopole case the same role that a pure gauge 2-form $B$ in the case of the ten dimensional KK-monopole. In this sense we will define the electric charge as associated with $C_{\m x^4}$ where the coordinate $x^4$ is the one of the $S^{1}$ fiber of the Hopf fibration. 

Let us stress that the main difference with the dyon effect for the ten dimensional KK-monopole is that we need to project the M-theory 3-form $C$ on the world-volume coordinates. Moreover, the fact that the so defined 2-form is a pure gauge is reflecting the gauge invariance of the world-volume BI lagrangian. 

Thus, in order to compute the electric charge, we parametrize the $V_\m$ as in \cite{Ima}

\be
V_\m=\left( f_1(r),0,\frac{1}{2}f_2(r)\cos\theta,f_2(r)\right),
\ee

\noindent which can be seen as a linear combination of the vierbeins (\ref{votn}). Therefore, using the fact that $f_2(r)$ is given in terms of the NUT potential $U$, we have the vector potential coming from $C_{\m\n}$ as

\be
\label{gffttf}
A^{(2)}=C_{\m x^4}=\frac{\a}{(r+4m)^2}dr.
\ee

\noindent where $\a$ is a dimensionfull constant which will be determined in the next section. Dualising this form we find

\be
\label{ef}
C_{(3)}={^*}C_{\m x^4}=3\a\left(\frac{r}{r+4m}\right)^2\sin\theta d\theta\wedge d\phi\wedge d\psi,
\ee

\noindent which is again proportional to the volume of the $S^3$. The integral of this form gives the (Julia-Zee) electric charge of the KK-monopole.

Notice that the $S^{1}$ part of the dyon moduli space is related to large gauge transformations of the 2-form $C_{\m\n}$.

%%%%%%%%%%%%%%%%%%%%%%%%%%%%%%%%%%%%%%%%%%%%%%%%%%%%%%%%%%%%%%%%%%%
%%%%%%%%%%%%%%%%%%%%%%%%%%%%%%%%%%%%%%%%%%%%%%%%%%%%%%%%%%%%%%%%%%%
\subsection{Reconstructing the moduli space}\label{ss66}

In this section we have argued that, as K-theory predicts, the D6-brane does have electric degrees of freedom which we have associated to the eleven dimensional 3-form. However, we have said nothing about the gauge group that SUGRA is seeing. To find it we can look for the transition functions defined by the vector fields $A^{(1)}$ and $A^{(2)}$ and their gauge transformations.

Mathematically, the moduli space of 't Hooft-Polyakov monopoles can be reconstructed as follows, \cite{Ati.Hit}. Consider the Hopf bundle $H$ over $S^2$. For a charge $k$ monopole, the direct sum $H^k\oplus H^{-k}$ defines another bundle over $S^2$ which can be extended radially over $\RR^3-\{ 0\}=S^2\times\RR^+$. This construction gives an $SU(2)$ bundle with transition functions

\be
k_{\a\b}:U_\a\cap U_\b\longrightarrow SU(2),
\ee

\noindent where

\be
\label{fts}
k_{\a\b}=\pmatrix{ g_{\a\b} & \cr
                            & h_{\a\b}},
\ee

\noindent and where

\be
\left( g_{\a\b}, h_{\a\b}\right):U_\a\cap U_\b \longrightarrow U(1),
\ee

\noindent are the transition functions for the (magnetic, electric) bundles. In ten dimensions, these bundles are the corresponding to the KK and H-monopoles respectively. In eleven dimensions, they correspond to the KK and C-monopole\footnote{we assing this name to the monopole coupled to the eleven dimensional 3-form.}.

In order to fix ideas, we need to get the right description of the differential forms involved, and the first step in this direction is the determinationi of the normalization constants. This can be done comparing the actions in eleven and ten dimensions. Let us begin with the bosonic part of the eleven dimensional supergravity

\be
S_{11}=\frac{1}{l_p^9}\int d^{11}x\sqrt{-g}\left( R-\frac{1}{48}\left( dC^{(3)}\right)^2+\frac{\sqrt{2}}{2^73^2}\frac{1}{\sqrt{-g}}C^{(3)}\wedge dC^{(3)}\wedge dC^{(3)}\right),
\ee

\noindent where, in $l_p$ units, the dimensions of the fields are

\be
\label{dimen}
\left[ dx\right]=\left[ d\right]=0,\quad \left[ R\right]=-2,\quad \left[ \sqrt{g}\right]=11,\quad \left[ g_{MN}\right]=2,\quad \left[ C^{(3)}\right]=3.
\ee

Let us firstly look at the magnetic part. In the dimensional reduction, we find the following term in the action

\be
S_{IIA}=\frac{1}{(2\pi)^7l_s^8}\int d^{10}x\dots-\frac{\sqrt{-g}}{4}\left( dA\right)^2,
\ee

\noindent from where it is not difficult to see that $[A]=1$.

On the other hand, the magnetic potential takes the form

\be
\label{pvm}
A_\m=k\frac{g_{\m\: 11}}{g_{11\: 11}}=k4m(1-\cos\theta)d\phi,
\ee

\noindent where $k$ is a dimensionless constant, as is obvious because $[m]=1$ and so already $[A]=1$. So there is just a numerical factor that we can fix to 1 with no lose of generality.

For the 3-form we proceed more carefully. On eleven dimensions, there is a coupling to the membrane of the form

\be
S_{M2}=\frac{g_{M2}}{3!}\int C^{(3)},
\ee

\noindent where, due to (\ref{dimen}), $[g_{M2}]=-3$, so we will take $g_{M2}\propto l_p^{-3}$. Now, because of the existence of the D6-brane, we have separated components for the 3-form, each one carrying different dimensions. We will impose $[A]=0$, and the relation between the other fields, equations (\ref{G}) and (\ref{F}) sets

\be
[B]=2,\quad [V]=1,\quad [C]=3.
\ee

\noindent By setting a trayectory for the wrapped membrane such that $(\theta,\phi)$ are constant, we can determine the dimensionfull constant $\a$ in which depends $C_{\m\n}$ as

\be
\a=\frac{3}{8\pi g_{M2}}.
\ee

The transition function for the $SU(2)$ bundle is then constructed from the $U(1)$ valued transition functions of these two complex line bundles as $(g_{\a\b},h_{\a\b})$, where $g_{\a\b}$ corresponds to the gauge transformation for (\ref{pvm}) and $h_{\a\b}$ were computed in (\ref{ftcb}). Taking into account the $1/\a'$ factor and the relation $m\propto l_s e^{\phi_0}$ and that a complex line bundle, $L$ satisfies $L^{-1}=L^*$, they can be written in $r\rightarrow\infty$ limit as

\be
\label{tffdsb}
(g_{\a\b},h_{\a\b})= (e^{i8m\phi}, e^{-i8m\phi}),
\ee

\noindent where $0\leq\phi\leq 2\pi$ is a polar coordinate in Taub-NUT. 

Therefore, from equation (\ref{tffdsb}) we conclude that the eleven dimensional monopole, or its ten dimensional counterpart, the D6-brane can be seen as a non-abelian monopole for a Yang-Mills-Higgs theory with $SU(2)$ gauge group. This implies that supergravity is not seeing the whole $U(2)$ K-theory group\footnote{see \cite{Gom.Man.3} for a discussion on this $U(1)$ factor.}. 

That this is indeed the case, we can use a spin from isospin construction (see \cite{Gom.Man.3} for more details), by embedding the spin connection (\ref{mf}), which has gauge group $SO(4)$, in one of its $SU(2)$ invariant subspaces. This can be done by means of the 't Hoot symbols \cite{THo.2}

\be
\bar\eta_i{}^{ab}=-\d^a{}_i\d^b{}_0+\d^a{}_0\d^b{}_i+\epsilon_i{}^{jk}\d^a{}_j\d^b{}_k,
\ee

\noindent where $i,j,\dots$ are $SU(2)$, the $\m,\n,\dots$ are space-time and $a,b,\dots$ are frame indices, so the gauge field is

\be
\label{escgg}
{\cal{A}}_\m^i=\frac{1}{2}\overline\eta^i{}_{ab}\omega_\m{}^{ab}.
\ee

Computing the components of the gauge field (\ref{escgg}), we find that 

\bea
\label{cecg}
{\cal{A}}^i{}_{x^4} & = & \frac{4m}{(r+4m)^2}dr, \\
\label{cmcg}
{\cal{A}}^i{}_\m & = & \left( 0,\d^i{}_3,\d^i{}_2\sin\theta\right)\frac{4m\left( K(r-4m)-1\right)}{r},
\eea

\noindent up to normalization, where

\be
K(r-4m)=\left(\frac{4m\left( 1-\cos\theta\right)}{r\sin\theta}\right)^2+2
\ee

In (\ref{cecg}) we see that the temporal component\footnote{this component is called the temporal one in an abuse of language. However it is the NUT coordinate and plays the same role in this gravitational set up as time in the field theoretical construction.} of the gauge field is precisely the field (\ref{gffttf}) and, therefore, this embedding reveals the eleven dimensional 3-form as the origin of the electric charge.

%%%%%%%%%%%%%%%%%%%%%%%%%%%%%%%%%%%%%%%%%%%%%%%%%%%%%%%%%%%%%%%%%%%%%%%%%%%%%%%%%%%%%%%%%%%%%%%%%%%%%%%%%%%%%%%%%%%%%%
%%%%%%%%%%%%%%%%%%%%%%%%%%%%%%%%%%%%%%%%%%%%%%%%%%%%%%%%%%%%%%%%%%%%%%%%%%%%%%%%%%%%%%%%%%%%%%%%%%%%%%%%%%%%%%%%%%%%%%
\section{Final remarks}\label{Con}

In this paper we have reviewed the classification of RR-charges in terms of K-theory mainly focusing in the effects of the B-field. We have seen that although different arguments say that this is the right classification, there are still certain unsolved subtleties. However, all of them seem to be surrounding the fact that there is a lack of a proper physical interpretation.

Focusing on the D6-brane, we have dealt with the deep relation between K-theory and eleven dimensions, given in terms of the eleven dimensional 3-form $C^{(3)}$. This allowed us to reveal the electric degrees of freedom of this soliton in terms of the dimensionally reduced 3-form. Reversing the argument, we can conclude that the electric degrees of freedom of the D6-brane reveal the existence of the eleventh (compact) dimension.

The relation between our results and the eleven dimensional 3-form can be related to those of \cite{Dia.Moo.Wit, e8}, where it was proposed that the topological sectors of the K-theory partition function in string theory and that of M-theory on the geometry $X\times S^1$ are equivalent if the eleven dimensional 3-form $C^{(3)}$ is a Chern-Simons form for an $E_8$ group. 

Concerning this point we have argued that the trace of an $E_8$ group in ten dimensions can be seen when we include the effects of non-torsion B-fields in the world-volume of a system of an infinite number of unstable D9-branes, in terms of its loop group $LE_8$, which is in accordance with previous arguments on this subject (see \cite{e8}).

Therefore we may conclude by saying that it seems that there is a heterotic origin of the closed string sector of string theory and that a proper physical interpretation of K-theory needs a precise interpretation of the closed string vacuum and its instabilities.

%%%%%%%%%%%%%%%%%%%%%%%%%%%%%%%%%%%%%%%%%%%%%%%%%%%%%%%%%%%%%%%%%%%%%%%%%%%%%%%%%%%%%%%%%%%%%%%%%%%%%%%%%%%%%%%%%%%%%%
%%%%%%%%%%%%%%%%%%%%%%%%%%%%%%%%%%%%%%%%%%%%%%%%%%%%%%%%%%%%%%%%%%%%%%%%%%%%%%%%%%%%%%%%%%%%%%%%%%%%%%%%%%%%%%%%%%%%%%
\section*{Acknowledgments}

I would like to thank N.~Alonso-Alberca, M.P.~Garc\'\i a del Moral, C.~G\'omez, N.~Hitchin, E.~L\'opez, P.~Meessen, P.~Resco, T.~Ort\'\i n, A.~Uranga and M.~Vila for useful conversations on topics related with this paper.

%%%%%%%%%%%%%%%%%%%%%%%%%%%%%%%%%%%%%%%%%%%%%%%%%%%%%%%%%%%%%%%%%%%%%%%%%%%%%%%%%%%%%%%%%%%%%%%%%%%%%%%%%%%%%%%%%%%%%%
%%%%%%%%%%%%%%%%%%%%%%%%%%%%%%%%%%%%%%%%%%%%%%%%%%%%%%%%%%%%%%%%%%%%%%%%%%%%%%%%%%%%%%%%%%%%%%%%%%%%%%%%%%%%%%%%%%%%%%

\bb

%%%%%%%%%%%%%%%%%%%%%%%%%%%%%%%%%%%%%%%%%%%%%%%%%%%%%%%%%%%%%%%%%%%%%%%%
%%%%%%%%%%%%%%%%%%%%%%%%%%% Introduction %%%%%%%%%%%%%%%%%%%%%%%%%%%%%%%

\bibitem{Pol} J.~Polchinski,
              {\em Dirichlet Branes and Ramond-Ramond charges},
              Phys.\ Rev.\ Lett.\ {\bf 75} (1995) 4724,
              {\tt hep-th/9510017}.

\bibitem{Sen.2} A.~Sen,
                {\em Stable non-BPS sates in string theory},
                J.\ High Energy Phys.\ {\bf 06} (1998) 007,
                {\tt hep-th/9803194}.\\
                ---,
                {\em Stable non-BPS bound states of BPS-branes},
                J.\ High Energy Phys.\ {\bf 08} (1998) 010,
                {\tt hep-th/9805019}.

\bibitem{Sen.1} A.~Sen,
               {\em Tachyon condensation on the brane anti-brane system},
               J.\ High Energy Phys.\  {\bf 08} (1998) 012,
               {\tt  hep-th/9805170}\\
               {\em Universality of the tachyon potential},
                J.\ High Energy Phys.\ {\bf 12} (1999) 027,
               {\tt  hep-th/9911116}.

\bibitem{Taq} J.~A.~Harvey, D.~Kutasov, E.~J.~Martinec,
              {\em On the relevance of tachyons},
              {\tt hep-th/0003101}.\\
              D.~Kutasov, M.~Mari\~no, G.~Moore,
              {\em Some exact results on tachyon condensation in string field theory},
              J.\ High Energy Phys.\ {\bf 0010} (2000) 045
              {\tt hep-th/0009148}.

\bibitem{Hor} P.~Ho\v rava,
               {\em Type IIA D-branes, K-theory and Matrix Theory},
               Adv.\ Theor.\ Math.\ Phys.\ {\bf 2} (1998) 1373-1404,
               {\tt hep-th/9812135}.

\bibitem{Min.Moo} R.~Minasian and G.~Moore,
                        {\em K-theory and Ramond-Ramond Charge},
                        J.\ High Energy Phys.\ {\bf 11} (1997) 002, 
                        {\tt hep-th/9710230}.

\bibitem{Wit.1} E.~Witten,
                  {\em D-branes and K-theory},
                  J.\ High Energy Phys.\ {\bf 12} (1998) 019,
                  {\tt hep-th/9810188}.

\bibitem{Moo.Wit} G.~Moore and E.~Witten,
                        {\em Self-duality, Ramond-Ramond Fields, and K-theory},
                        J.\ High Energy Phys.\ {\bf 05} (2000) 032,
                        {\tt hep-th/9912279}.

\bibitem{Wit.2} E.~Witten,
                {\em Bound states of strings and p-branes},
                Nucl.\ Phys.\ {\bf B460} (1996) 335-350,
                {\tt hep-th/9510135}

\bibitem{Gree.Har.Moo} M.B.~Green, J.A.~Harvey and G.~Moore,
                      {\em I-brane inflow and anomalous couplings on D-branes},
                      Class.\ Quant.\ Grav.\ {\bf 14} (1997) 47,
                      {\tt hep-th/9605033}.

\bibitem{Che.Yin} E.~Cheung and Z.~Yin,
                  {\em Anomalies, branes and currents},
                  Nucl.\ Phys.\ {\bf B517} (1998) 69,
                  {\tt hep-th/9710230}.

\bibitem{Shift} E.~Witten,
                {\em Duality relations among topological effects in string theory},
                J.\ High Energy Phys.\ {\bf 05}, (200) 044,
                {\tt hep-th/9912279}.\\

\bibitem{Fre.Wit} D.~Freed and E.~Witten,
                   {\em Anomalies in String Theory with D-branes},
                   {\tt hep-th/9907189}.

\bibitem{Gom} C.~G\'omez,
                        {\em A Comment on the K-theory Meaning of the Topology Gauge Fixing},
                        {\tt hep-th/0104211}.

\bibitem{THo.1} G.~'t Hooft,
                {\em Topology of the gauge condition and new confinement phases in non-abelian gauge theories},
                Nucl.\ Phys.\ {\bf B190} (1981) 455.

\bibitem{Ati.Bot.Sha} M.F.~Atiyah, R.~Bott and A.~Shapiro,
                      {\em Clifford modules},
                      Topology {\bf 1} (1962) 245.

\bibitem{Ros} J.~Rosenberg,
              {\em Continuous-trace Algebras from the Bundle Theoretical Point of View},
              J.\ Austral.\ Math.\ Soc.\ (Series A) {\bf 47} (1989) 368-381.

\bibitem{Don.Kar} P.~Donovan and M.~Karoubi,
                  {\em Graded Brauer groups and K-theory with local coefficients},
                  IHES Pub.\ {\bf 38} (1970) 5.

\bibitem{Kap} A.~Kapustin,
              {\em D-branes in a topologically nontrivial B-field},
              Adv.\ Theor.\ Math.\ Phys.\ {\bf 4} (2000) 127-154,
              {\tt hep-th/9909089}.

\bibitem{Sei.Wit} N.~Seiberg and E.~Witten,
                  {\em String theory and noncommutative geometry},
                  J.\ High Energy Phys.\ {\bf 09} (1999) 032,
                  {\tt hep-th/9908142}.

\bibitem{Bou.Mat} P.~Bouwknegt and V.~Mathai,
                  {\em D-branes, B-fields and Twisted K-theory},
                  J.\ High Energy Phys.\ {\bf 03} (2000) 007,
                  {\tt hep-th/0002023}.

\bibitem{Ati} M.F.~Atiyah,
             {\em K-theory},
             {\tt (W.A. Benjamin, New York, 1967)}

\bibitem{Kar} M.~Karoubi, 
              {\em K-theory, an introduction},
              {\tt (Springer-Verlag, 1978)}

\bibitem{Wit.4} E.~Witten,
                {\em An overview of K-theory applied to strings},
                Int.\ J.\ Mod.\ Phys.\ {\bf A16} (2001) 693-706. Also in *Ann Arbor 2000, Strings* 53-66,
                {\tt hep-th/0007175}

\bibitem{Sza} R.~Szabo,
              {\em D-branes, tachyons and K-homology},
              Mod.\ Phys.\ Lett.\ {\bf A17} (2002) 2297-2316,
              {\tt hep-th/0209210}.

\bibitem{Gom.Man.3} C.~G\'omez and J.J.~Manjar\'\i n, 
                    {\em to appear}.

\bibitem{Mat.Sin} V.~Mathai and I.M.~Singer,
                  {\em Twisted K-homology theory, twisted {\sl {Ext}}-Theory},
                  {\tt hep-th/0012046}.

\bibitem{Per.1} V.~Periwall,
                {\em D-brane charges and K-homology},
                J.\ High Energy Phys.\ {\bf 07} (2000) 041,
                {\tt hep-th/0006223}.

\bibitem{Gom.Man.2} C.~G\'omez and J.J.~Manjar\'\i n,
                  {\em Dyons, K-theory and M-theory},
                  {\tt hep-th/0111169}.

\bibitem{Gom.Man.1} C.~G\'omez and J.J.~Manjar\'\i n,
                  {\em A note on the dyonic D6-brane},
                  {\sl to appear in Int.\ J.\ Mod.\ Phys.\ {\bf A}}, Contribution to the 6th International Workshop on Conformal Field Theory and Integrable Models, Landau Institute Sept. 2002,
                  {\tt hep-th/0302096}.

%%%%%%%%%%%%%%%%%%%%%%%%%%%%%%%%%%%%%%%%%%%%%%%%%%%%%%%%%%%%%%%%%%%%%%%%
%%%%%%%%%%%%%%%%%%%%% K-theory in string theory %%%%%%%%%%%%%%%%%%%%%%%%

\bibitem{Ols.Sza} K.~Olsen and R.J.~Szabo,
                      {\em Constructing D-branes from K-theory},
                      {\tt hep-th/9907140}.

\bibitem{Hul}C.~Hull,
                        {\em Gravitational duality, branes and charges},
                        Nucl.\ Phys.\ {\bf B509} (1998) 216-251,
                        {\tt  hep-th/9705162}.

\bibitem{Gre} M.B.~Green,
              {\em Point-like states for type IIB string theory},
              Phys.\ Lett.\ {\bf B329} (1994) 435,
              {\tt hep-th/9403040}.

\bibitem{Per.2} V.~Periwal,
                {\em Antibranes and corssing symmetry},
                {\tt  hep-th/9612215}.

\bibitem{Qui} D.~Quillen,
              {\em Superconnections and the Chern character},
              Topology {\bf 24} (1985) 89.

\bibitem{Sre} M.~Srednicki,
              {\em IIB or not IIB},
              J.\ High Energy Phys.\ {\bf 08} (1998) 005,
              {\tt hep-th/9807138}.

\bibitem{Osc} O.~Garc\'\i a-Prada,
              {\em Seiberg-Witten invariants and vortex equations},
              in {\tt Sym\'etries quantiques. Quantum symmetries, Les Houches, session LXIV, 1995}. A.~Connes, K.~Gawedzki and J.~Zinn-Justin eds.(Elseivier Science, 1998).

\bibitem{Wit.3} E.~Witten,
               {\em Global gravitational anomalies},
               Comm.\ Math.\ Phys. {\bf 100} (1985) 197-229.

\bibitem{Bis.Fre} J.M.~Bismut and D.~Freed,
                  {\em The analysis of elliptic families: dirac operators, eta invariants, and the holonomy theorem of Witten},
                  Comm.\ Math.\ Phys. {\bf 107} (1986) 103-163.

\bibitem{Fre} D.~Freed,
              {\em Determinants, torsion, and strings},
              Comm.\ Math.\ Phys. {\bf 107} (1986) 483-513.

\bibitem{Ati.Pat.Sin} M.F.~Atiyah, V.K.~Patodi and I.M.~Singer,
                      {\em Spectral symmetry and Riemannian geometry. I},
                      Math.\ Proc.\ Cambridge Philos.\ Soc.\ {\bf 77} (1975) 43-69.
                      ---,---,---,
                      {\em Spectral symmetry and Riemannian geometry. II},
                      Math.\ Proc.\ Cambridge Philos.\ Soc.\ {\bf 78} (1975) 405-432.
                      ---,---,---,
                      {\em Spectral symmetry and Riemannian geometry. III},
                      Math.\ Proc.\ Cambridge Philos.\ Soc.\ {\bf 79} (1976) 71-99.

%%%%%%%%%%%%%%%%%%%%%%%%%%%%%%%%%%%%%%%%%%%%%%%%%%%%%%%%%%%%%%%%%%%%%%%
%%%%%%%%%%%%%%%%%%%%%% Non-trivial B-fields %%%%%%%%%%%%%%%%%%%%%%%%%%%

\bibitem{Con} A.~Connes,
               {\em Noncommutative geometry},
               {\tt Academic Press, 1994}.

\bibitem{Cal.Har.Str} C.G.~Callan,Jr., J.A.~Harvey and A.~Strominger,
                      {\em Supersymmetric string solitons}, 
                     In *Trieste 1991, Proceedings, String theory and quantum gravity '91* 208-244 and Chicago Univ. - EFI 91-066 (91/11,rec.Feb.92)
                     {\tt hep-th/9112030}.

\bibitem{Nek.Sch} N.~Nekrasov and A.~Schwarz,
                      {\em Instantons on noncommutative $\mathbb{R}^4$ and $(2,0)$ superconformal six dimensional theory},
                      {\tt hep-th/9802068}.

\bibitem{Ati.Hit} M.F.~Atiyah and N.J.~Hitchin,
                      {\em The geometry and dynamics of magnetic monopoles},
                      {\tt Princeton Univ.\ Press}, Princeton, NJ, 1988.

\bibitem{Nek.Dou} N.~Nekrasov and M.~Douglas,
                  {\em Noncommutative field theory},
                  Rev.\ Mod.\ Phys.\ {\bf 73} (2001) 977-1029,
                  {\tt hep-th/0106048}

\bibitem{Kon} M.~Kontsevich,
              {\em Deformation quatization of Poisson manifolds I},
              {\tt q-alg/9709040}.

\bibitem{Cor.Sch} L.~Cornalba and R.~Schiappa,
                  {\em Nonassociative star product deformations for D-brane worldvolumes in curved backgrounds},
                  Commun.\ Math.\ Phys.\ {\bf 225} (2002) 33-66,
                  {\tt hep-th/0101219}.

\bibitem{Orl} O.~\'Alvarez,
              {\em Topological Quantization and Cohomology},
              Commun.\ Math.\ Phys.\ {\bf 100} (1985) 279.

\bibitem{Hit} N.~Hitchin,
              {\em Lectures on Special Lagrangian Submanifolds},
              Lectures given both at the Winter School on Mirror Symmetry in Harvard, January 1999 and the school on differential geometry held at ICTP, Trieste in April 1999.
              {\tt math.DG/9907034}.

\bibitem{Sen.3} A.Sen,
              {\em Kaluza-Klein Dyons in String Theory},
              Phys.\ Rev.\ Lett.\ {\bf 79} (1997) 1619-1621, 
              {\tt hep-th/9705212}.

\bibitem{Greg.Har.Moo} R.~Gregory, J.A.~Harvey and G.~Moore,
              {\em Unwinding Strings and T-duality of Kaluza-Klein and H-Monopoles},
              Adv.\ Theor.\ Math.\ Phys.\ {\bf 1} (1997) 283-297,
              {\tt hep-th/9708086}.

\bibitem{Ima} Y.~Imamura,
                             {\em Born-Infeld Action and Chern-Simons Term from Kaluza-Klein Monopole in M-theory},
                             Phys.\ Lett.\ {\bf B414} (1997) 242-250,
                             {\tt hep-th/9706144}.

%%%%%%%%%%%%%%%%%%%%%%%%%%%%%%%%%%%%%%%%%%%%%%%%%%%%%%%%%%%%%%%%%%%%%%
%%%%%%%%%%%%%%%%%%%%%% K-theory with B-fields %%%%%%%%%%%%%%%%%%%%%%%%

\bibitem{Jan} K.~J\"anich,
             {\em Vektorraumb\"undel und der Raum der Fredholm-Operatoren},
             Math.\ Ann.\ {\bf 161} (1965) 129-142.

\bibitem{Ati.Sin} M.I.~Atiyah and I.M.~Singer,
                  {\em Index Theory for Skew-Adjoint Fredholm Operators},
                  I.H.E.S.\ Publ.\ Math.\ {\bf 37} (1969) 305-326.

\bibitem{Boo.Ble} B.~Booss and D.D.~Bleecker,
                  {\em Topology and Analysis: The Atiyah-Singer Index Formula and Gauge-Theoretic Physics},
                  {\tt (Springer-Verlag, 1977)}.\\
                  ---,---,
                  {\em Spectral Invariants of Operators of Dirac Type on Partitioned Manifolds},
                  {\tt http://mmf.ruc.dk/~booss/ell/lbo030307p.pdf}

\bibitem{Dix.Dou} J.~Dixmier and A.~Douady,
                  {\em Champs continus d'spaces Hilbertiens et de C$^*$-alg\`ebres},
                  Bull.\ Soc.\ Math.\ France {\bf 91} (1963) 227-284.

\bibitem{Gel.Nai} I.M.~Gelfand and M.A.~Na\u\i mark,
                  {\em On the Embeding of Normed Rings into the Ring of Operators in Hilbert Space},
                  Mat.\ Sbornik {\bf 12} (1943) 197-213.

\bibitem{Gro} A.~Grothendieck,
              {\em Le grupe de Brauer I},
              S\'eminaire Bourbaki, Exp. No. {\bf 290} (1964/1965) 01-21.

\bibitem{Wit.9} E.Witten,
                {\em Baryons and branes in anti-de Sitter space},
                J.\ High Eenergy Phys.\ {\bf 07} (1998) 006,
                {\tt hep-th/9805112}.

\bibitem{Mal.Moo.Sei} J.~Maldacena, G.~Moore and N.~Seiberg,
                          {\em D-brane instantons and K-theory charges},
                          J.\ High Energy Phys.\ {\bf 11} (2001) 062,
                          {\tt hep-th/ 0108100}

\bibitem{Moo} G.~Moore,
              {\em K-theory from a physical perspective},
              Talk given at Symposium on Topology, Geometry and Quantum Field Theory, Oxford, England, United Kingdom, 24-29 Jun 2002.
              {\tt hep-th/0304018}

\bibitem{Gop.Min.Str} R.~Gopakumar, S.~Minwalla and A.~Strominger,
                      {\em Noncommutative solitons},
                      J.\ High Eenergy Phys.\ {\bf 05} (2000) 020,
                      {\tt hep-th/0003160}.

\bibitem{Das.Muk.Raj} K.~Dasgupta, S.~Mukhi and G.~Rajesh,
                      {\em Noncommutative tachyons},
                      J.\ High Energy Phys.\ {\bf 06} (2000) 022,
                      {\tt hep-th/0005006}.

\bibitem{Har.Moo} J.A.~Harvey and G.~Moore,
                 {\em Noncommutative tachyons and K-theory},
                 J.\ Math.\ Phys.\ {\bf 42} (2001) 2765-2780,
                 {\tt hep-th/0009030}.

\bibitem{Wit.st} E.~Witten,
                 {\em Non-commutative geometry and string field theory},
                 Nucl.\ Phys.\ {\bf B268} (1986) 253.

\bibitem{Ber.1} N.~Berkovits,
                {\em Review of open superstring field theory},
                {\tt hep-th/0105230}.

\bibitem{Wit.5} E.~Witten,
                {\em Noncommutative tachyons and string field theory},
                {\tt hep-th/0006071}.

\bibitem{Kui} N.H.~Kuiper,
              {\em The homotopy type of the unitary group of Hilbert space},
              Topology {\bf 3} (1965) 19.

\bibitem{Pal} R.S.~Palais,
              {\em On the homotopy type of certain groups of operators},
              Topology {\bf 3} (1965) 271.

\bibitem{Har} J.~Harvey,
              {\em Topology of the gauge group in noncommutative gauge theory},
              To appear in the proceedings of Strings 2001: International Conference, Mumbai, India, 5-10 Jan 2001,
              {\tt hep-th/0105242}.

\bibitem{Asa.Sug.Ter} T.~Asakawa, S.~Sugimoto and S.Terashima,
                      {\em D-branes, matrix theory and K-homology},
                      J.\ High Energy Phys.\  {\bf 03} (2002) 034,
                      {\tt hep-th/0108085}.

\bibitem{Van} I.V.~Vancea,
              {\em On the algebraic K-theory of the massive D8 and M9 branes},
              Int.\ J.\ Mod.\ Phys.\ {\bf A16} (2001) 4429-4452,
              {\tt hep-th/9905034}.

\bibitem{Ber.Jej.Lei} D.~Berenstein, V.~Jejjala, R.G.~Leigh
                     {\em Marginal and relevant deformations of N=4 field theories and noncommutative moduli spaces of vacua}
                     Nucl.\ Phys.\ {\bf B589} (2000) 196-248,
                     {\tt hep-th/0005087}.

\bibitem{Lec.Pop.Sza} O.~Lechtenfeld, A.D.~Popov and R.J.~Szabo,
                      {\em Noncommutative instantons in higher dimensions, vortices and topological K-cycles},
                      J.\ High Eenergy Phys.\ {\bf 12} (2003) 022,
                      {\tt hep-th/0310267}.

\bibitem{Dia.Moo.Wit} E.~Diaconescu, G.~Moore and E.~Witten,
               {\em $E_8$ Gauge Theory, and a Derivation of K-theory from M-theory},
               {\tt hep-th/0005090}. 
               ---, ---, ---,
               {\em A Derivation of K-theory from M-theory},
               {\tt hep-th/0005091}.

%%%%%%%%%%%%%%%%%%%%%%%%%%%%%%%%%%%%%%%%%%%%%%%%%%%%%%%%%%%%%%%%%%%%%%%%%%%%%%%%%%%%%%
%%%%%%%%%%%%%%%%%%%%%%%%%%%%%%%%% The D6-brane %%%%%%%%%%%%%%%%%%%%%%%%%%%%%%%%%%%%%%%

\bibitem{e8} A.~Adams and J.~Evslin,
             {\em The loop group of E(8) and K-theory form 11D},
             J.\ High Energy Phys.\ {\bf 02} (2003) 029,
             {\tt hep-th/0203218}.\\
             J.~Evslin and H.Sati,
             {\em Susy versus E(8) gauge theory in eleven dimensions}
             J.\ High Energy Phys.\ {\bf 05} (2003) 048,
             {\tt hep-th/0210090}.\\
             J.~Evslin,
             {\em From E(8) to F via T},
             {\tt hep-th/0311235}.\\
             V.~Mathai and H.~Sati,
             {\em Some relations between twisted K-theory and E(8) gauge theory},
             J. High Energy Phys.\ {\bf 03} (2004) 016,
             {\tt hep-th/0312033}.\\
             E.~Diaconescu, G.~Moore and D.~Freed,
             {\em The M-theory three form and E(8) gauge theory} 
             {\tt hep-th/0312069}.

\bibitem{dyon}  J.S.~Schwinger,
                {\em A magnetic model of matter},
                Science {\bf 165} (1969) 757-761,\\
                D.~Zwanziger,
                {\em Quantum field theory of particles with both electric and magnetic charges},
                Phys.\ Rev.\ {\bf 176} (1968) 1489-1495.\\
                B.~Julia and A.~Zee,
                {\em Poles with both magnetic and electric charges in non-Abelian gauge theory},
                Phys.\ Rev.\ {\bf D11} (1975) 2227-2232.

\bibitem{dira} P.A.M.~Dirac,
               {\em The theory of magnetic monopoles},
               Phys.\ Rev.\ {\bf 74} (1948) 817-830.

\bibitem{thpol} G.~'t Hooft,
                {\em  Magnetic monopoles in unified theories},
                Nucl.\ Phys.\ {\bf B79} (1974) 276.\\
                A.M.~Polyakov,
                {\em Particle spectrum in the quantum field theory},
                JETP Lett.\ {\bf 20} (1974) 194.

\bibitem{bps}  M.K.~Prasad and M.~Sommerfield,
               {\em An exact classical solution for the 't Hooft monopole and the Julia-Zee dyon},
               Phys.\ Rev.\ Lett.\ {\bf 35} (1975) 760-762.\\
               E.B.~Bogomol'nyi,
               {\em Stability of classical solutions},
               Sov.\ J.\ Nucl.\ Phys.\ {\bf 24} (1976) 449.

\bibitem{Wit.6} E.~Witten,
               {\em Dyons of charge $\frac{e\theta}{2\pi}$. },
               Phys.\ Lett.\ {\bf B86} (1979) 283-287.

\bibitem{Gib.Rub} G.W.~Gibbons and P.J.~Ruback,
                    {\em Winding Strings, Kaluza-Klein Monopoles and Runge-Lenz Vectors},
                    Phys.\ Lett.\ {\bf B215} (1988) 653-656.

\bibitem{Sor} R.~Sorkin,
              {\em Kaluza-Klein Monopole},
              Phys.\ Rev.\ Lett.\ {\bf 51} 2 (1983) 87

\bibitem{Gro.Per} D.J.~Gross and M.J.~Perry,
             {\em Magnetic Monopoles in Kaluza-Klein Theories},
             Nucl. Phys. {\bf B226} (1983) 29.

\bibitem{Gib.Haw} G.~Gibbons and S.~Hawking,
             {\em Classification of gravitational instanton symmetries},
             Comm.\ Math.\ Phys.\  {\bf 66} (1979) 291-310.

\bibitem{Mal.Moo.Sei.1} J.~Maldacena, G.~Moore and N.~Seiberg,
                        {\em D-brane charges in five-brane backgrounds},
                        J.\ High Energy Phys.\ {\bf 10} (2001) 005,
                        {\tt hep-th/0108152}.

\bibitem{Gre.Sch} M.~Green and J.H.~Schwarz,
             {\em Anomaly Cancellation in Supersymmetric $D=10$ Gauge Theory and Superstring Theory},
             Phys.\ Lett.\ {\bf B149} (1984) 117-122.

\bibitem{Hol.Pro} J.W.~Van Holten and A.~Van Proeyen,
                  {\em  N=1 supersymmetry algebras in D=2, D=3, D=4 mod-8},
                  J.\ Phys.\ Gen.\ {\bf A15} (1982) 3763.

\bibitem{Tow} P.~Townsend,
              {\em The eleven dimensional membrane revisited},
              Phys.\ Lett.\ {\bf B350} (1995) 184-187,
              {\tt  hep-th/9501068}.\\
              ---,
              {\em M-theory from its superalgebra},
              {\tt hep-th/9712004}.

\bibitem{Sen.4} A.~Sen,
                {\em Dynamics of multiple Kaluza-Klein monopoles in M and string theory},
                Adv.\ Theor.\ Math.\ Phys.\ {\bf 1} (1998) 115-126,
                {\tt hep-th/9707042}.

\bibitem{Ber.Ort.Jan} E.~Bergshoeff, B.~Janssen and T.~Ortin,
                           {\em Kaluza-Klein Monopoles and Gauged Sigma Models},
                            Phys.\ Lett.\ {\bf B410} (1997) 131-141,
                            {\tt hep-th/9706117}.

\bibitem{Spa} J.~Sparks,
              {\em Global world sheet anomalies from M-theory}
              {\tt hep-th/0310147}.

\bibitem{Hor.Wit} P.~Ho\v rava and E.~Witten,
             {\em Heterotic And Type I String Dynamics From Eleven Dimensions},
             Nucl.\ Phys.\ {\bf B460} (1996) 506,
             {\tt hep-th/9510209};\\
             {\em Eleven-Dimensional Supergravity On A Manifold With Boundary},
             Nucl.\ Phys.\ {\bf B475} (1996) 94-114,
             {\tt hep-th/9603142}.

\bibitem{Wit.7} E.~Witten, 
               {\em On Flux Quatization in M-theory and the effective action},
               J.\ Geom.\ Phys.\ {\bf 22} (1997) 1-13,
               {\tt hep-th/9609122}.

%%%%%%%%%%%%%%%%%%%%%%%%%%%%%%%%%%%%%%%%%%%%%%%%%%%%%%%%%%%%%%%%%%%%%%%%%%%%%%%%%%%%%%
%%%%%%%%%%%%%%%%%%%%%%%%%%%%%%% Moduli space of monopoles %%%%%%%%%%%%%%%%%%%%%%%%%%%%

\bibitem{naka} H.~Nakajima,
                          {\em Heisenberg algebra and Hilbert schemes of points on porjective surfaces},
                          {\tt alg-geom/9507012};\\
                          {\em Lectures on Hilbert schemes of points on surfaces}, 
                          {\tt University Lecture series, Vol 18, American Mathematical Society, 1999}.

\bibitem{adhm} M.F.~Atiyah, V.G.~Drinfeld, N.J.~Hitchin and Yu I.~Manin,
                           {\em Construction of instantons},
                           Phys.\ Lett.\ {\bf A65} (1978) 185-187.

\bibitem{THo.2} G.~'t Hooft,
                {\em Symmetry breaking through Bell-Jackiw anomalies},
                Phys.\ Rev.\ Lett.\ {\bf 37} (1976) 8-11.

\eb

\end{document}